\documentclass[aps,prl,twocolumn,amsmath,amssymbs,superscriptaddress,longbibliography]{revtex4-1}
\usepackage{epsfig}
\usepackage{wrapfig}
\usepackage{bbm}
\usepackage[usenames]{color}
\usepackage{array}
\usepackage{times}

\usepackage{graphicx}
\usepackage{float}
\usepackage{multirow}
\usepackage{natbib,twoopt}
\usepackage[breaklinks=true]{hyperref}
\usepackage[dvipsnames]{xcolor}
\usepackage{amsmath,amsfonts,amssymb}
\usepackage{graphicx}
\usepackage{hyperref}
\usepackage{color}
\usepackage{stackengine}
\usepackage{subfigure}
\usepackage{verbatim}
\usepackage{stmaryrd} 


\usepackage{soul}

\newcommand{\RNum}[1]{\uppercase\expandafter{\romannumeral #1\relax}}

\newcommand{\balancecolsandclearpage}{%
	\close@column@grid
	\cleardoublepage
	\twocolumngrid
}

\begin{document}



\title{Observation of the magic angle and flat band physics in dipolar photonic lattices}

\author{Diego Rom\'an-Cort\'es}
\email{These authors have equally contributed to this work}
\affiliation{Departamento de Física and Millenium Institute for Research in Optics–MIRO, Facultad de Ciencias Físicas y Matemáticas, Universidad de Chile, 8370448 Santiago, Chile}

\author{Maxim Mazanov}
\email{These authors have equally contributed to this work}
\affiliation{School of Physics and Engineering, ITMO University, Saint  Petersburg 197101, Russia}

\author{Rodrigo A. Vicencio}
\email{rvicencio@uchile.cl}
\affiliation{Departamento de Física and Millenium Institute for Research in Optics–MIRO, Facultad de Ciencias Físicas y Matemáticas, Universidad de Chile, 8370448 Santiago, Chile}

\author{Maxim A. Gorlach}
\email{m.gorlach@metalab.ifmo.ru}
\affiliation{School of Physics and Engineering, ITMO University, Saint  Petersburg 197101, Russia}

\begin{abstract}
Evanescently coupled waveguide arrays provide a tabletop platform to realize a variety of Hamiltonians, where physical waveguides correspond to the individual sites of a tight-binding lattice. Nontrivial spatial structure of the waveguide modes enriches this picture and uncovers further possibilities. Here, we demonstrate that the effective coupling between $p$-like modes of adjacent photonic waveguides changes its sign depending on their relative orientation vanishing for a proper alignment at a so-called \textit{magic angle}. Using femtosecond laser-written waveguides, we demonstrate this experimentally for $p$-mode dimers and graphene-like photonic lattices exhibiting quasi-flat bands at this angle. We observe diffraction-free propagation of corner and bulk states providing a robust experimental evidence of a two-dimensional Aharonov-Bohm-like caging in an optically switchable system.
\end{abstract}

\maketitle

\section{Introduction}

Photonic lattices have emerged as a versatile platform to explore localization and transport phenomena~\cite{Szameit2010,Rechtsman2013}. As the paraxial equation governing light propagation in a single waveguide resembles Schr{\"o}dinger equation in quantum mechanics, arrays of evanescently coupled waveguides provide a straightforward realization of paradigmatic tight-binding models~\cite{Haus1987Jan} with the possibility to accomodate such exotic features as negative couplings~\cite{Keil2016,Caceres-Aravena2022Jun}, artificial gauge fields~\cite{Rechtsman2013,Mukherjee_PRL} and non-Hermitian physics~\cite{Eichelkraut2014,Weimann2016,vicencioNR}.

\begin{figure*}[ht!]
	\centering
	\includegraphics[width=0.8\textwidth]{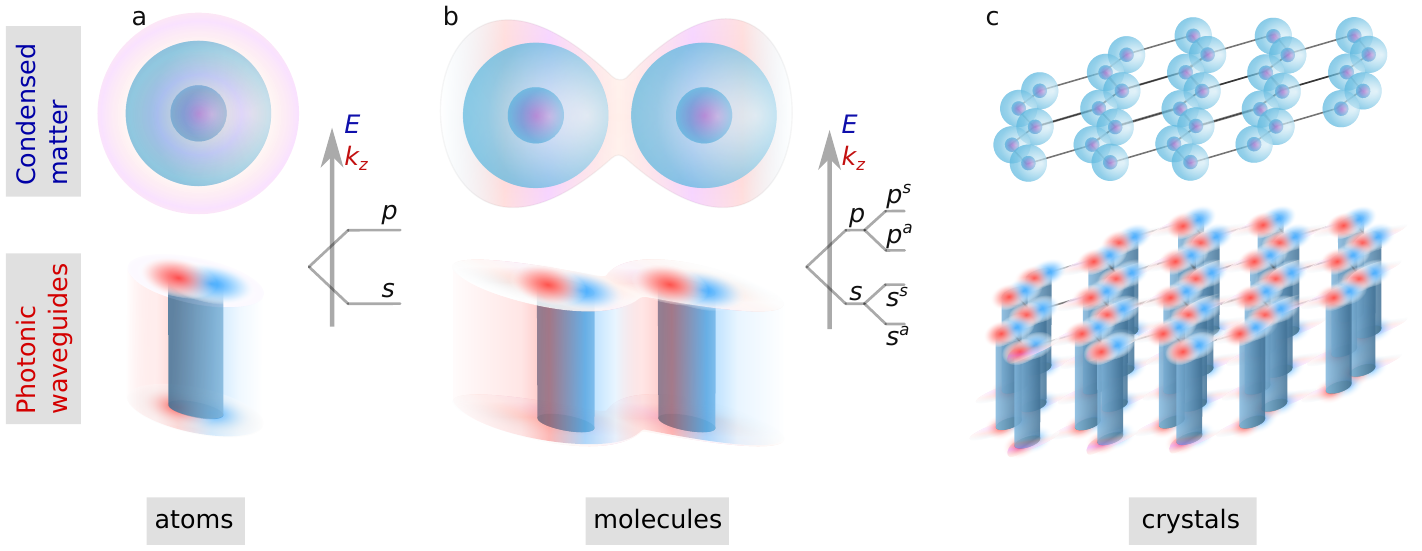}
        \caption{ 
        \textbf{Parallels between condensed matter systems and photonic lattices}: 
        \textbf{a}~individual atoms, \textbf{b}~molecules and \textbf{c}~2D crystals. Classical light propagation in waveguide arrays models temporal evolution of condensed matter systems. 
        Second row illustrates $p$-mode waveguides~(\textbf{a}) as promising artificial-atom building blocks which could be utilized in photonic molecules with nontrivial phase profiles~(\textbf{b}), and flat-band photonic lattices~(\textbf{c}). Zero coupling between the $p$ modes at a magic angle decomposes the modes of a $p$-mode graphene-like lattice into isolated compact localized states rendering all bands flat.} 
        \label{fig:F0}
\end{figure*}


The development of this platform exhibits profound parallels with condensed matter physics and material science~\cite{Longhi2009Apr} as illustrated in Fig.~\ref{fig:F0}. On a qualitative level, the formation of a material can be sketched as a coupling of atoms and formation of molecules; hybridization of molecular orbitals results in molecular aggregates and, finally, in the emergence of a 2D material. In the same spirit, the waveguides can be strongly coupled to each other to form photonic molecules~\cite{mol1us}. The collective modes of those molecules with distinct symmetry can be readily identified as different orbitals, and their overlap determines the properties of a given photonic lattice. This parallel appears to be especially fruitful in light of the recent studies which demonstrated photonic molecule concept experimentally~\cite{mol1us} and predicted a vast range of novel functionalities, including exotic topological phases mediated by the coupling between different photonic orbitals~\cite{PRASP1D,SavelevGorlach_PRB,Schulz2022Nov,Mazanov2022}.

Here, we add a crucial ingredient to this picture by introducing a {\it magic angle} in photonic lattices. The idea of a magic angle has previously materialized in condensed matter literature~\cite{Cao2018}, where it was introduced to denote unconventional electronic properties of twisted bilayer graphene, with flat bands and resulting superconducting behavior~\cite{Heikki,Torma1}. The  advancement of this concept resulted in the field of twistronics~\cite{Ciarrocchi2022} spanning such disciplines as condensed matter physics, polaritonics and  photonics~\cite{Hu2021,Hua_LPR_2020}. Adding to the versatility of photonic lattices, here we experimentally demonstrate that a suitable arrangement of optical waveguides hosting $p$-type orbital modes results in a vanishing effective coupling between them. As a result, the lattice of such waveguides features dispersionless bulk modes unlocking an all-bands-flat regime (ABF)~\cite{Danieli2021Aug1,CaceresPRL2024}. A distinctive feature of this regime is photonic Aharonov-Bohm (AB) caging when an arbitrary initial state does not diffract into the bulk of the lattice remaining confined in a finite spatial region, oscillating and periodically recovering its initial profile.

The origin of flat bands can be traced to the so-called compact localized states~\cite{Flach2014,Maimaiti2017Mar,Maimaiti2019Mar,Maimaiti2021Apr}, which are perfectly confined lattice eigenmodes. Typically, flat bands in tight-binding lattices arise due to the destructive interference in the so-called connector sites~\cite{FBluis,Leykam_APX,Leykam2018Jul,vicencioReportFB} and require fine tuning of the lattice parameters~\cite{finetuningFB}. However, having an {\it all} band flat regime is much less trivial. While in one-dimensional systems ABF regime could be achieved by several means including multi-orbital mechanisms~\cite{PRASP1D,SavelevGorlach_PRB,Caceres-Aravena2022Jun,Jorg2020Aug,Pelegri2019Feb,MazanovIEEE} or effective magnetic fields~\cite{Vidal_PRL,Longhi_OL,Mukherjee_PRL,Caceres-Aravena2022Jun,CaceresPRL2024}, its realization in two dimensions is challenging, with a few theoretical proposals~\cite{Vidal_PRL,MazanovIEEE} and a single experiment~\cite{Yang2024Feb} utilizing inhomogeneous lattices. Hence, an experimental realization of ABF regime in periodic two-dimensional lattices remains a fundamental open problem in Physics.


In this Article, we realize all-bands-flat regime experimentally by harnessing magic angle paradigm. We start from a pair of interacting $p$-mode waveguides identifying the conditions for a zero coupling between them at a magic angle. Next, we arrange the waveguides into a graphene-like lattice and predict the formation of quasi-flat bands enabling several cycles of Aharonov-Bohm oscillations. By imaging the intensity distribution at the output, we demonstrate a diffraction-suppressed propagation of bulk and topological corner modes of the designed lattice.

\section{Results}

%

\begin{figure*}[t!]
	\centering
	\includegraphics[width=0.99\textwidth]{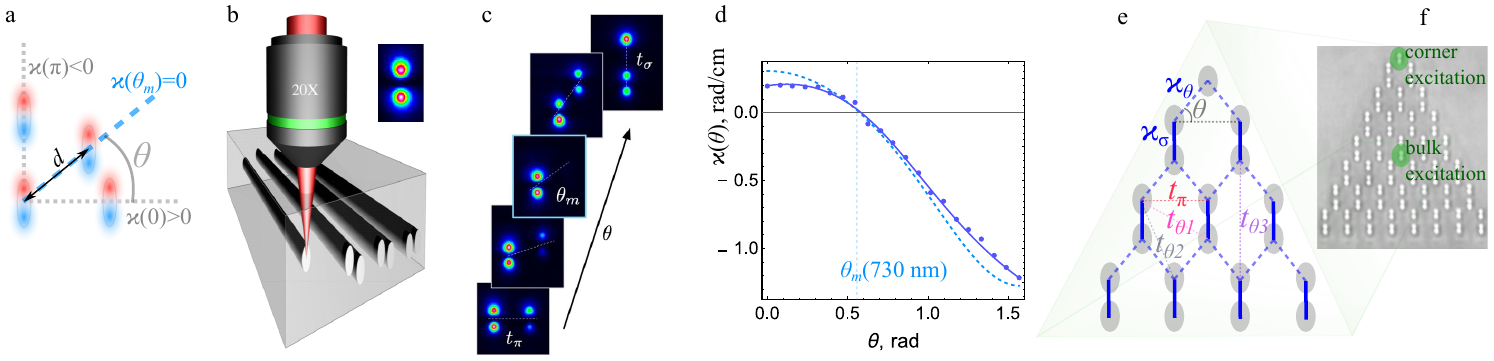}
	\caption{ 
        \textbf{The magic angle concept and $p$-mode graphene-like lattice.} 
        \textbf{a} Coupling between the pair of identical $p$-mode optical waveguides vanishes at a magic angle $\theta_m$. 
        \textbf{b} Illustration of the femtosecond laser writing technique for the waveguide dimers. 
        The inset shows an experimental $p$-mode intensity profile after excitation via the spatial light modulator (SLM)~\cite{Supplement}, at the working wavelength $730$~nm. 
        \textbf{c} Dimer intensity profiles for the propagation distance of $L = 25$~mm after an SLM excitation at the left waveguide. 
        \textbf{d} Experimentally extracted $p$-mode coupling $\kappa(\theta)$ (blue dots, with solid line as a guide to the eye) and numerically calculated coupling from collective dimer propagation constants (dashed line). 
        \textbf{e} Sketch of the graphene-like lattice with fixed nearest-neighbour distance $d$, two dominant couplings $\varkappa_\sigma$ and $\varkappa_\theta$ indicated by solid and dashed lines, respectively, as well as long-range couplings $t_\pi$ and $t_{\theta1}-t_{\theta3}$. 
        \textbf{f} White light microscope image of a fabricated lattice with bulk and corner excitation spots highlighted in green. 
        }
        \label{fig:F1}
\end{figure*}

{\bf Magic angle for $p$-mode waveguides.}~\label{sec:1} 
In case of two parallel electric dipoles, simple electrostatic calculation suggests that their interaction vanishes at an angle $\theta_0=\text{arcsin}\left(1/\sqrt{3}\right)\approx 0.615$ when the dipoles are perfectly decoupled. Though this idea has been appreciated in chemistry~\cite{chemicalMA18,Kazmaier1994Oct}, it breaks down at nonzero frequencies. In contrast, magic angle in photonics [Fig.~\ref{fig:F1}\textbf{a}] persists in a wide frequency range. As the coupling between the two evanescent waveguide modes is determined by their overlap integral~\cite{yariv_optical_2003,vicencioNR}, the signs of the coupling can be read off from Fig.~\ref{fig:F1}\textbf{a}. For horizontal positioning ($\theta=0$), both pairs of the closest lobes are in phase resulting in positive coupling, while for vertical positioning ($\theta=\pi/2$) the overlap integral is negative dominated by the interaction of the two opposite-phase lobes. Since the coupling varies continuously, a special angle $\theta_m$ exists such that coupling $\kappa(\theta_m)=0$.


To probe the interaction between the $p$ modes experimentally, we fabricate pairs of elliptical waveguides with vertically oriented $p$ modes utilizing a femtosecond laser writing technique~\cite{Szameit2010,Supplement},  Fig.~\ref{fig:F1}\textbf{b}. Launching light in one of the waveguides, we track intensity beatings in the dimer [Fig.~\ref{fig:F1}\textbf{c}] retrieving the absolute value of the effective coupling constant. The coupling sign is chosen based on the continuity of $\kappa(\theta)$ function and assuming $\kappa(0)>0$, $\kappa(\pi/2)<0$.

In particular, experiments with $p$ dimers at a distance $d=25\ \mu$m and wavelength $\lambda = 730$~nm yield magic angle $\theta_m\approx 0.56$ rad, not much different from the one predicted in electrostatics. Experimentally retrieved values of the coupling constant $\kappa$ for various $\theta$ are depicted in Fig.~\ref{fig:F1}\textbf{d}. Numerical calculations in COMSOL Multiphysics, with refractive index $n = 1.48$ (borosilicate), elliptical waveguide profiles with a fine-tuned contrast of $\delta n = 10^{-3}$, semi-axes $a=2.45\,\mu$m and $b=8.18\,\mu$m~\cite{Supplement} fully support these findings as shown by the dotted line in Fig.~\ref{fig:F1}\textbf{d}.

\begin{figure*}[ht!]
	\centering
	\includegraphics[width=0.99\textwidth]{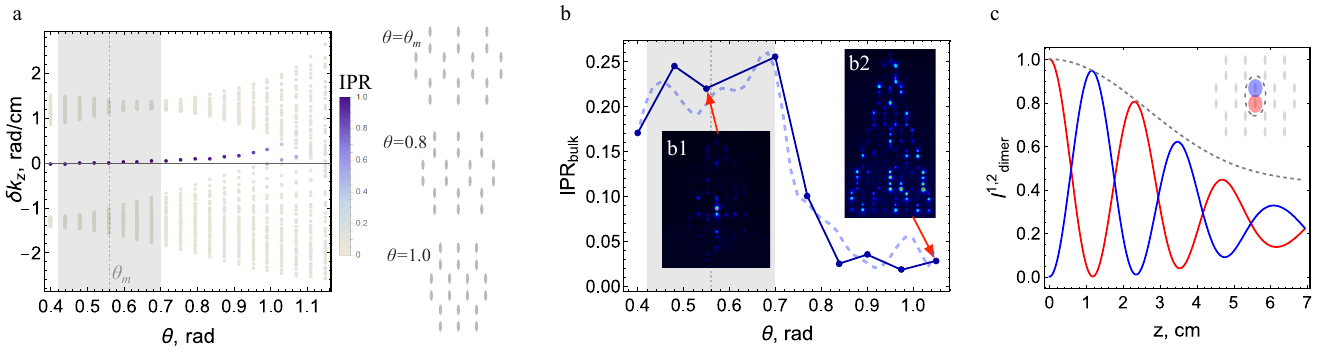}
	\caption{ 
        \textbf{Bulk flat bands and photonic Aharonov-Bohm caging.}
        \textbf{a} Spectrum of the propagation constants for a finite $15$-row graphene-like lattice calculated for numerically retrieved couplings depending on angle $\theta$ for $d = 25\ \mu$m and $\lambda = 730$ nm. Color encodes the localization of the modes. 
        This includes the effect of long-range couplings and the nonorthogonality correction. Right insets show the geometry of the lattice for several representative $\theta$.
        \textbf{b} Values of inverse participation ratio (IPR) versus angle $\theta$ for a bulk site SLM excitation. Dotted line shows simulated results, dots with solid approximating line depict experimental data. We observe a plateau behaviour with nearly isolated dimer dynamics at angles around the magic angle $\theta_m \simeq 0.56$ (gray shaded area). Insets show experimental propagated intensity profiles for two angles highlighted by arrows. 
        \textbf{c} Numerically propagated intensities for $\theta = \theta_m$ after bulk excitation (red, its dimer partner in blue) show an oscillating behavior indicative of photonic caging, with almost three full caging cycles at the lattice output facet. Gray dashed line shows the total dimer intensity.
        }
        \label{fig:F2}
\end{figure*}

\begin{figure*}[t!]
	\centering
	\includegraphics[width=0.99\textwidth]{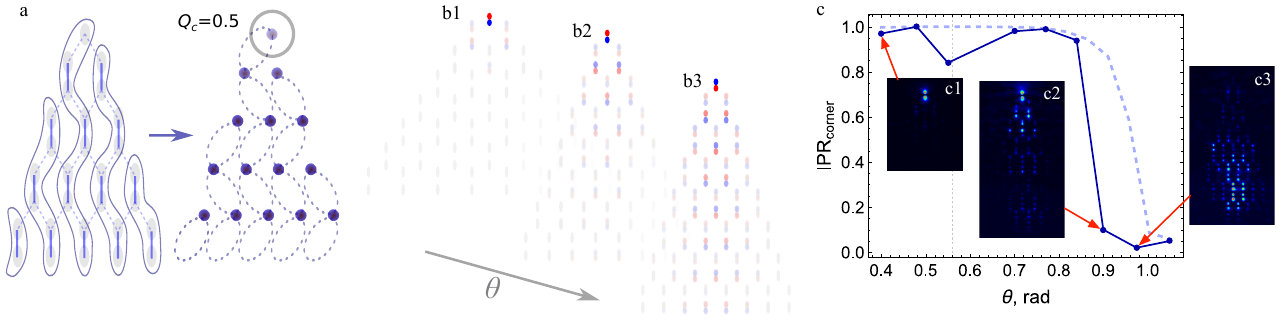}
	\caption{ 
        \textbf{Topological corner state.} 
        \textbf{a}~Wannier centers following from the flat-band limit decomposition of the lattice into SSH arrays of isolated dimers indicate the effective corner charge $Q_c = 0.5$ and quantized bulk polarization $P = (0.5, 0.5)$. 
        \textbf{b1-b3}~Amplitude profiles of the topological corner state in three finite lattices for the wavelength $\lambda = 730$~nm and three angles $\theta_{1-3} = 0.4, 0.9, 0.975$~rad. 
        \textbf{c} Inverse participation ratio for the output field demonstrates localization transition at large angles $\theta > 0.9$~rad. 
         Insets depict the experimental intensity profiles for geometries \textbf{b1}--\textbf{b3}, respectively.
        }
        \label{fig:F3}
\end{figure*}


\begin{figure*}[t!]
	\centering
	\includegraphics[width=0.99\textwidth]{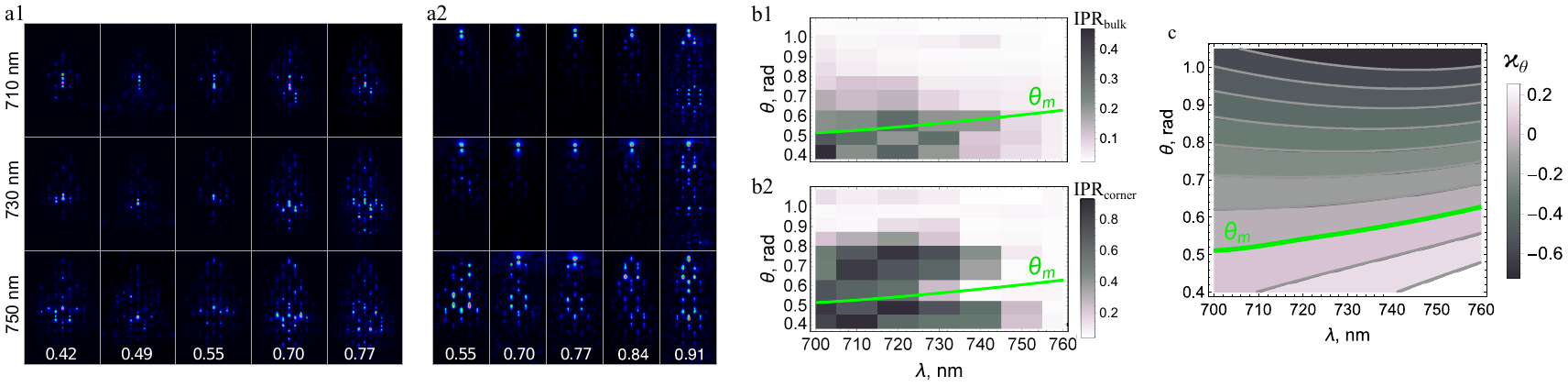}
	\caption{ 
        \textbf{Wavelength dependence of caging and corner-localized dynamics.}
        \textbf{a} Output intensity profiles for bulk (\textbf{a1}) and corner (\textbf{a2}) site laser excitations for a selected set of wavelengths and angles (indicated). 
        \textbf{b} IPR for the bulk (\textbf{b1}) and corner (\textbf{b2}) site excitations obtained from the propagated intensity profiles, with the numerical magic angle dependence shown by a green curve. 
        \textbf{c} Numerical nearest-neighbour $p$-$p$ coupling $\kappa_\theta$ as a function of the wavelength and the angle $\theta$, showing a larger magic angle at larger wavelengths.}
        \label{fig:F4}
\end{figure*}

{\bf Optical graphene-like lattice.}~\label{sec:1} 
In turn, magic angle physics has a profound effect on the behavior of $p$ waveguide optical lattices. To illustrate that, we design and fabricate a set of 9 graphene-like structures consisting of identical $p$-mode waveguides with a fixed distance $d$ between the nearest neighbors [Figs.~\ref{fig:F1}\textbf{e},\textbf{f}]. The properties of the lattices are then controlled by the angle $\theta$ defining the degree of geometric deformation and varying from sample to sample [Fig.~\ref{fig:F1}\textbf{e}]. Qualitatively, if coupling $\kappa_\theta$ vanishes, the lattice decomposes into isolated dimers with two fully flat bulk bands formed by symmetric and anti-symmetric dimer modes. The isolated upper corner site in a finite triangular sample forms a ``zero-energy'' corner-state in the middle of the bandgap. 



Realistically, photonic bands do not become fully flat. Although the nearest-neighbor coupling $\kappa_\theta$ is tuned to be zero at the magic angle, long-range couplings $t_\pi$ and $t_{\theta 1}$-$t_{\theta 3}$ have to be taken into account as well. We thus construct a tight-binding Hamiltonian of the form 
%
\begin{eqnarray}
\label{fullH}
    \hat{H}_{\Sigma} = 
    \hat{H}_{NN} + \hat{H}_{NNN} + \hat{H}_{NNNN}\ , 
\end{eqnarray}
%
where $\hat{H}_{NN}$ includes two major nearest-neighbour 
couplings $\kappa_\theta$ and $\kappa_\sigma$, $\hat{H}_{NNN}$~-- two next-nearest-neighbour couplings $t_\pi$ and $t_{\theta2}$, and $\hat{H}_{NNNN}$~-- two longer-range couplings $t_{\theta1}$ and $t_{\theta3}$, as depicted in Fig.~\ref{fig:F1}\textbf{e} (see details in Supplemental Materials~\cite{Supplement}).  

Furthermore, the modes of isolated waveguides which are used as a basis for the tight-binding description, are not quite orthogonal, as there is a nonzero overlap between them quantified by the coefficients
\begin{equation}
    c_{i j}
    =
    \frac{1}{I} \iint E_j(x, y) E_i^{*}(x, y) d x d y \
 , 
\end{equation}
where $I = \iint | E_i(x, y) |^2 d x d y=\iint | E_j(x, y) |^2 d x d y$ and $E_{i,j}$ are the electric field profiles at waveguides $i$ and $j$, respectively. Note that non-orthogonality corrections become especially pronounced for small ($\theta <0.4$ rad) and large ($\theta > 0.9$ rad) angles resulting in a modified propagation equation~\cite{Haus1987Jan,Maczewsky,Schulz2022Mar}
\begin{equation}
    - i \frac{d \boldsymbol{\psi}(z)}{d z}
    =
    \hat{c}^{-1} \hat{H}_{\Sigma} \boldsymbol{\psi}(z)\ 
    , 
\end{equation}
where $\boldsymbol{\psi}(z)$ is a vector constructed from the amplitudes of the waveguide fields.


Though the matrix $\hat{H}'_{\Sigma} = \hat{c}^{-1} \hat{H}_{\Sigma}$ is non-Hermitian, it can be brought to the  Hermitian form by the suitable non-unitary linear transformation to a new orthogonalized basis~\cite{Schulz2022Mar}. As a consequence, the entire dynamics is Hermitian and the spectrum of propagation constants is real. The latter can be recovered both from the Hermitian Hamiltonian in the transformed basis or the initial matrix $\hat{H}'_{\Sigma}$.


The results computed for a finite lattice, taking into account both long-range couplings and non-orthogonality corrections, are presented in Fig.~\ref{fig:F2}\textbf{a}. We observe that the spectrum indeed contains nearly flat bands persisting in the entire region of angles $\theta$ highlighted in Fig.~\ref{fig:F2}\textbf{a} by gray.

Along with the nearly flat bands, we find   a localized corner state near the center of the bandgap with high inverse participation ratio~\cite{Thouless_1974}
$\text{IPR}=\sum_{i} I_{i}^4/\left(\sum_{i} I_{i}^2\right)^2 \approx 1$. Here, $I_i$ is the intensity at $i^{\text{th}}$ site and the sum runs over all sites of the lattice. Notably, the corner mode arises in the middle of the bandgap in the same $\theta$ interval, shifting away from the bandgap center for larger angles.
%


{\bf Bulk dynamics and Aharonov-Bohm caging.}~\label{sec:1} 
We probe the flat-band dynamics experimentally by exciting a single bulk site [see Fig.~\ref{fig:F1}\textbf{f}] on a set of 9 graphene-like lattices fabricated on a $L = 70$~mm long glass wafer~\cite{Supplement}. In this case, we use a spatial light modulator setup~\cite{Supplement} and track the propagated intensity profiles at the output, as shown in Fig.~\ref{fig:F2}\textbf{b}. We then post-process the data to extract the inverse participation ratio (IPR) quantifying the localization of the field: higher IPR indicates stronger field confinement.

The analyzed data shows two distinct regimes: for lattices with $0.4<\theta<0.8$ close to the magic angle a good caging effect is observed. However, larger angles ($\theta>0.8$) favor a significant field diffraction, so the field at the output is delocalized. These experimental results shown in Fig.~\ref{fig:F2}{\bf b} by the dots are in a very good agreement with the theoretical calculation (dashed line).
Thus, contrary to the expectation that the long-range couplings and non-orthogonality corrections could degrade the flat-band dynamics~\cite{LiebLeykam,Maczewsky}, in our lattice they do not destroy quasi-flat band physics around the magic angle.

%

Numerical simulations of the field propagation for  excited dimer near the magic angle show a coherent and slowly decaying oscillating behavior which is characteristic of caging effects in the presence of small band dispersion, with nearly three full cycles of caging dynamics shown in Fig.~\ref{fig:F2}\textbf{c}. This indicates similar or even increased caging fidelity compared to the recent 1D photonic proposals~\cite{Mukherjee_PRL,Jorg2020Aug,Caceres-Aravena2022Jun}. The excited region collapses to dimer compact localized states (CLS), being therefore the smallest observed flat band state in 2D lattices~\cite{Liebus,LiebSeba,vicencioReportFB}, similar to CLS observed in 1D systems~\cite{SebaOL}. This further confirms the quasi-flat band dynamics arising due to the magic angle effect. 

{\bf Corner state excitation.}~\label{sec:2} 
Another alternative is to excite the corner site of the fabricated graphene-like lattice [Fig.~\ref{fig:F1}\textbf{e}], which is expected to decouple from the rest of the structure in the idealized magic angle limit. In this case, the lattice decomposes into isolated dimers, which could be further grouped into vertical quasi-one-dimensional Su-Schrieffer-Heeger~(SSH) arrays~\cite{1DSSH}, as described in Fig.~\ref{fig:F3}\textbf{a}. Note that only one of those chains has a nontrivial charge $Q_c = 0.5$ at the edge, quantized by the inversion symmetry~\cite{Quantization_2019}. 

This atomic limit is adiabatically connected to all realistic configurations which still respect inversion symmetry, and thus the corner charge persists for the full Hamiltonian $\hat{H}_{\Sigma}$. A direct calculation of Wannier centers~\cite{Supplement} yields bulk polarization $P = (0.5,0.5)$, quantized by the inversion symmetry. Together with the corner charge, this allows us to classify the corner mode as a Wannier-type higher-order topological one in the $h_{1b}^{(4)}$ primitive generator class for $C_2$-symmetric systems~\cite{Quantization_2019}. Finite-lattice calculations utilizing full Hamiltonian~\eqref{fullH} confirm the existence of the corner state in the entire interval of angles $\theta$, with stronger localization arising close to the magic angle, Fig.~\ref{fig:F3}\textbf{b}. 
%


In our experiments, we excite the corner $p$-waveguide using spatial light modulator (SLM) image setup. We then image the field distribution at the output facet and extract the inverse participation ratio (IPR) quantifying the localization of the field. The results for the different lattices (i.e. different angles $\theta$) are presented in Fig.~\ref{fig:F3}\textbf{c}. These results suggest that the IPR experiences a rapid decrease at angles $\theta>0.8$, signaling less effective excitation of the corner mode. In this region of $\theta$, the corner mode becomes less localized developing a broader tail which degrades the efficiency of a single-site excitation.



{\bf Wavelength dependence of the magic angle.}~\label{sec:3} 
Interestingly, magic angle physics and the designed quasi-flat band lattice can be readily reconfigured by changing the excitation wavelength. In our experiments, we use free-space single site excitation from a supercontinuum laser source~\cite{Supplement} in the wavelength range $\lambda\in\{700,760\}$ nm for 7 representative wavelengths. The obtained results depicted  in Fig.~\ref{fig:F4}{\bf a1-a2} suggest that the flat-band caging dynamics and the corner state localization are both preserved up to a wavelength $\lambda<750$~nm. At larger wavelengths, the modes become more extended and the influence of long-range couplings deteriorates the overall caging effect. 

The retrieved localization (IPR) of the output field for various $\theta$ and wavelenghts is color-coded in  Figs.~\ref{fig:F4}\textbf{b1-b2} both for the bulk and corner excitations, indicating the presence of stronger caging in darker regions of the plot. We observe a gradual growth of the magic angle with the excitation wavelength, which aligns with numerical simulations Figs.~\ref{fig:F4}\textbf{c} and opens a route towards photonic caging tunable by the excitation wavelength. A similar trend in the magic angle is observed when the distance $d$ between the waveguides is decreased~\cite{Supplement}.



\section{Discussion and conclusions}\label{sec:Discussion}


%
%

In summary, we observed experimentally and confirmed numerically the magic angle phenomenon for $p$-mode photonic waveguides, evidencing a complete cancellation of coupling and an effective invisibility for a particular waveguide alignment. Importantly, for a given geometrical configuration the near-magic-angle $p$-mode waveguide could be fine-tuned by changing the operating wavelength providing a switchable coupler with all-optical control over the coupling efficiency. 

The $p$-mode graphene-like lattice possesses a nearly all-bands-flat phase with a highly localized and relatively long-lasting caging effect. To the best of our knowledge, our observation constitutes the first demonstration of such behaviour in periodic and linear two-dimensional lattice systems serving as an alternative to the celebrated condensed-matter proposal of a Dice lattice in the magnetic field~\cite{Vidal_PRL}, which is challenging to implement in current photonic systems. 

The formation of a higher-order topological corner-localized state was clearly observed for the entire range of lattice geometries, providing an additional propagation channel which is partially protected by the on-site $p$-mode symmetries. Both caging phenomenon and corner mode dynamics are most pronounced for shorter wavelengths, where the influence of long-range couplings and mode non-orthogonality is less important. 

We anticipate that the similar magic angle phenomenon arises for higher-order orbital modes, transforming the magic angle idea into a broad physical concept. Such multipolar modes could be utilized to suppress not only nearest neighbor, but also next-nearest neighbor couplings enabling the systematic construction of flat bands in electromagnetic, acoustic and condensed matter systems and consolidating our observation as a beginning of orbital physics engineering. 


\section*{Acknowledgments}
 
Theoretical models describing flat band formation were supported by the Russian Science Foundation (grant No.~24-72-10069). Numerical studies of waveguide coupling were supported by Priority 2030 Federal Academic Leadership Program. M.M. and M.A.G. acknowledge partial support from the Foundation for the Advancement of Theoretical Physics and Mathematics ``Basis''. Experimental and other numerical studies were supported by Millennium Science Initiative Program ICN17$\_$012 and FONDECYT Grant 1231313.




\bibliography{refs}

\begin{thebibliography}{55}%
\makeatletter
\providecommand \@ifxundefined [1]{%
 \@ifx{#1\undefined}
}%
\providecommand \@ifnum [1]{%
 \ifnum #1\expandafter \@firstoftwo
 \else \expandafter \@secondoftwo
 \fi
}%
\providecommand \@ifx [1]{%
 \ifx #1\expandafter \@firstoftwo
 \else \expandafter \@secondoftwo
 \fi
}%
\providecommand \natexlab [1]{#1}%
\providecommand \enquote  [1]{``#1''}%
\providecommand \bibnamefont  [1]{#1}%
\providecommand \bibfnamefont [1]{#1}%
\providecommand \citenamefont [1]{#1}%
\providecommand \href@noop [0]{\@secondoftwo}%
\providecommand \href [0]{\begingroup \@sanitize@url \@href}%
\providecommand \@href[1]{\@@startlink{#1}\@@href}%
\providecommand \@@href[1]{\endgroup#1\@@endlink}%
\providecommand \@sanitize@url [0]{\catcode `\\12\catcode `\$12\catcode `\&12\catcode `\#12\catcode `\^12\catcode `\_12\catcode `\%12\relax}%
\providecommand \@@startlink[1]{}%
\providecommand \@@endlink[0]{}%
\providecommand \url  [0]{\begingroup\@sanitize@url \@url }%
\providecommand \@url [1]{\endgroup\@href {#1}{\urlprefix }}%
\providecommand \urlprefix  [0]{URL }%
\providecommand \Eprint [0]{\href }%
\providecommand \doibase [0]{http://dx.doi.org/}%
\providecommand \selectlanguage [0]{\@gobble}%
\providecommand \bibinfo  [0]{\@secondoftwo}%
\providecommand \bibfield  [0]{\@secondoftwo}%
\providecommand \translation [1]{[#1]}%
\providecommand \BibitemOpen [0]{}%
\providecommand \bibitemStop [0]{}%
\providecommand \bibitemNoStop [0]{.\EOS\space}%
\providecommand \EOS [0]{\spacefactor3000\relax}%
\providecommand \BibitemShut  [1]{\csname bibitem#1\endcsname}%
\let\auto@bib@innerbib\@empty
\bibitem [{\citenamefont {Szameit}\ and\ \citenamefont {Nolte}(2010)}]{Szameit2010}%
  \BibitemOpen
  \bibfield  {author} {\bibinfo {author} {\bibfnamefont {Alexander}\ \bibnamefont {Szameit}}\ and\ \bibinfo {author} {\bibfnamefont {Stefan}\ \bibnamefont {Nolte}},\ }\bibfield  {title} {\enquote {\bibinfo {title} {Discrete optics in femtosecond-laser-written photonic structures},}\ }\href {\doibase 10.1088/0953-4075/43/16/163001} {\bibfield  {journal} {\bibinfo  {journal} {Journal of Physics B: Atomic, Molecular and Optical Physics}\ }\textbf {\bibinfo {volume} {43}},\ \bibinfo {pages} {163001} (\bibinfo {year} {2010})}\BibitemShut {NoStop}%
\bibitem [{\citenamefont {Rechtsman}\ \emph {et~al.}(2013)\citenamefont {Rechtsman}, \citenamefont {Zeuner}, \citenamefont {Plotnik}, \citenamefont {Lumer}, \citenamefont {Podolsky}, \citenamefont {Dreisow}, \citenamefont {Nolte}, \citenamefont {Segev},\ and\ \citenamefont {Szameit}}]{Rechtsman2013}%
  \BibitemOpen
  \bibfield  {author} {\bibinfo {author} {\bibfnamefont {Mikael~C.}\ \bibnamefont {Rechtsman}}, \bibinfo {author} {\bibfnamefont {Julia~M.}\ \bibnamefont {Zeuner}}, \bibinfo {author} {\bibfnamefont {Yonatan}\ \bibnamefont {Plotnik}}, \bibinfo {author} {\bibfnamefont {Yaakov}\ \bibnamefont {Lumer}}, \bibinfo {author} {\bibfnamefont {Daniel}\ \bibnamefont {Podolsky}}, \bibinfo {author} {\bibfnamefont {Felix}\ \bibnamefont {Dreisow}}, \bibinfo {author} {\bibfnamefont {Stefan}\ \bibnamefont {Nolte}}, \bibinfo {author} {\bibfnamefont {Mordechai}\ \bibnamefont {Segev}}, \ and\ \bibinfo {author} {\bibfnamefont {Alexander}\ \bibnamefont {Szameit}},\ }\bibfield  {title} {\enquote {\bibinfo {title} {{Photonic Floquet topological insulators}},}\ }\href {\doibase 10.1038/nature12066} {\bibfield  {journal} {\bibinfo  {journal} {Nature}\ }\textbf {\bibinfo {volume} {496}},\ \bibinfo {pages} {196--200} (\bibinfo {year} {2013})}\BibitemShut {NoStop}%
\bibitem [{\citenamefont {Szameit}\ and\ \citenamefont {Rechtsman}(2024)}]{Szameit2024}%
  \BibitemOpen
  \bibfield  {author} {\bibinfo {author} {\bibfnamefont {Alexander}\ \bibnamefont {Szameit}}\ and\ \bibinfo {author} {\bibfnamefont {Mikael~C.}\ \bibnamefont {Rechtsman}},\ }\bibfield  {title} {\enquote {\bibinfo {title} {Discrete nonlinear topological photonics},}\ }\href {\doibase 10.1038/s41567-024-02454-8} {\bibfield  {journal} {\bibinfo  {journal} {Nature Physics}\ }\textbf {\bibinfo {volume} {20}},\ \bibinfo {pages} {905--912} (\bibinfo {year} {2024})}\BibitemShut {NoStop}%
\bibitem [{\citenamefont {Haus}\ \emph {et~al.}(1987)\citenamefont {Haus}, \citenamefont {Huang}, \citenamefont {Kawakami},\ and\ \citenamefont {Whitaker}}]{Haus1987Jan}%
  \BibitemOpen
  \bibfield  {author} {\bibinfo {author} {\bibfnamefont {H.}~\bibnamefont {Haus}}, \bibinfo {author} {\bibfnamefont {W.}~\bibnamefont {Huang}}, \bibinfo {author} {\bibfnamefont {S.}~\bibnamefont {Kawakami}}, \ and\ \bibinfo {author} {\bibfnamefont {N.}~\bibnamefont {Whitaker}},\ }\bibfield  {title} {\enquote {\bibinfo {title} {{Coupled-mode theory of optical waveguides}},}\ }\href {\doibase 10.1109/JLT.1987.1075416} {\bibfield  {journal} {\bibinfo  {journal} {J. Lightwave Technol.}\ }\textbf {\bibinfo {volume} {5}},\ \bibinfo {pages} {16--23} (\bibinfo {year} {1987})}\BibitemShut {NoStop}%
\bibitem [{\citenamefont {Keil}\ \emph {et~al.}(2016)\citenamefont {Keil}, \citenamefont {Poli}, \citenamefont {Heinrich}, \citenamefont {Arkinstall}, \citenamefont {Weihs}, \citenamefont {Schomerus},\ and\ \citenamefont {Szameit}}]{Keil2016}%
  \BibitemOpen
  \bibfield  {author} {\bibinfo {author} {\bibfnamefont {Robert}\ \bibnamefont {Keil}}, \bibinfo {author} {\bibfnamefont {Charles}\ \bibnamefont {Poli}}, \bibinfo {author} {\bibfnamefont {Matthias}\ \bibnamefont {Heinrich}}, \bibinfo {author} {\bibfnamefont {Jake}\ \bibnamefont {Arkinstall}}, \bibinfo {author} {\bibfnamefont {Gregor}\ \bibnamefont {Weihs}}, \bibinfo {author} {\bibfnamefont {Henning}\ \bibnamefont {Schomerus}}, \ and\ \bibinfo {author} {\bibfnamefont {Alexander}\ \bibnamefont {Szameit}},\ }\bibfield  {title} {\enquote {\bibinfo {title} {{Universal Sign Control of Coupling in Tight-Binding Lattices}},}\ }\href {\doibase 10.1103/physrevlett.116.213901} {\bibfield  {journal} {\bibinfo  {journal} {Physical Review Letters}\ }\textbf {\bibinfo {volume} {116}},\ \bibinfo {pages} {213901} (\bibinfo {year} {2016})}\BibitemShut {NoStop}%
\bibitem [{\citenamefont {C{\ifmmode\acute{a}\else\'{a}\fi}ceres-Aravena}\ \emph {et~al.}(2022)\citenamefont {C{\ifmmode\acute{a}\else\'{a}\fi}ceres-Aravena}, \citenamefont {Guzm{\ifmmode\acute{a}\else\'{a}\fi}n-Silva}, \citenamefont {Salinas},\ and\ \citenamefont {Vicencio}}]{Caceres-Aravena2022Jun}%
  \BibitemOpen
  \bibfield  {author} {\bibinfo {author} {\bibfnamefont {Gabriel}\ \bibnamefont {C{\ifmmode\acute{a}\else\'{a}\fi}ceres-Aravena}}, \bibinfo {author} {\bibfnamefont {Diego}\ \bibnamefont {Guzm{\ifmmode\acute{a}\else\'{a}\fi}n-Silva}}, \bibinfo {author} {\bibfnamefont {Ignacio}\ \bibnamefont {Salinas}}, \ and\ \bibinfo {author} {\bibfnamefont {Rodrigo~A.}\ \bibnamefont {Vicencio}},\ }\bibfield  {title} {\enquote {\bibinfo {title} {{Controlled Transport Based on Multiorbital Aharonov-Bohm Photonic Caging}},}\ }\href {\doibase 10.1103/PhysRevLett.128.256602} {\bibfield  {journal} {\bibinfo  {journal} {Phys. Rev. Lett.}\ }\textbf {\bibinfo {volume} {128}},\ \bibinfo {pages} {256602} (\bibinfo {year} {2022})}\BibitemShut {NoStop}%
\bibitem [{\citenamefont {Mukherjee}\ \emph {et~al.}(2018)\citenamefont {Mukherjee}, \citenamefont {Di~Liberto}, \citenamefont {\"Ohberg}, \citenamefont {Thomson},\ and\ \citenamefont {Goldman}}]{Mukherjee_PRL}%
  \BibitemOpen
  \bibfield  {author} {\bibinfo {author} {\bibfnamefont {Sebabrata}\ \bibnamefont {Mukherjee}}, \bibinfo {author} {\bibfnamefont {Marco}\ \bibnamefont {Di~Liberto}}, \bibinfo {author} {\bibfnamefont {Patrik}\ \bibnamefont {\"Ohberg}}, \bibinfo {author} {\bibfnamefont {Robert~R.}\ \bibnamefont {Thomson}}, \ and\ \bibinfo {author} {\bibfnamefont {Nathan}\ \bibnamefont {Goldman}},\ }\bibfield  {title} {\enquote {\bibinfo {title} {{Experimental Observation of {A}haronov-{B}ohm Cages in Photonic Lattices}},}\ }\href {\doibase https://doi.org/10.1103/PhysRevLett.121.075502} {\bibfield  {journal} {\bibinfo  {journal} {Physical Review Letters}\ }\textbf {\bibinfo {volume} {121}},\ \bibinfo {pages} {075502} (\bibinfo {year} {2018})}\BibitemShut {NoStop}%
\bibitem [{\citenamefont {Eichelkraut}\ \emph {et~al.}(2014)\citenamefont {Eichelkraut}, \citenamefont {Weimann}, \citenamefont {St{\"u}tzer}, \citenamefont {Nolte},\ and\ \citenamefont {Szameit}}]{Eichelkraut2014}%
  \BibitemOpen
  \bibfield  {author} {\bibinfo {author} {\bibfnamefont {T.}~\bibnamefont {Eichelkraut}}, \bibinfo {author} {\bibfnamefont {S.}~\bibnamefont {Weimann}}, \bibinfo {author} {\bibfnamefont {S.}~\bibnamefont {St{\"u}tzer}}, \bibinfo {author} {\bibfnamefont {S.}~\bibnamefont {Nolte}}, \ and\ \bibinfo {author} {\bibfnamefont {A.}~\bibnamefont {Szameit}},\ }\bibfield  {title} {\enquote {\bibinfo {title} {Radiation-loss management in modulated waveguides},}\ }\href {\doibase 10.1364/ol.39.006831} {\bibfield  {journal} {\bibinfo  {journal} {Optics Letters}\ }\textbf {\bibinfo {volume} {39}},\ \bibinfo {pages} {6831} (\bibinfo {year} {2014})}\BibitemShut {NoStop}%
\bibitem [{\citenamefont {Weimann}\ \emph {et~al.}(2016)\citenamefont {Weimann}, \citenamefont {Kremer}, \citenamefont {Plotnik}, \citenamefont {Lumer}, \citenamefont {Nolte}, \citenamefont {Makris}, \citenamefont {Segev}, \citenamefont {Rechtsman},\ and\ \citenamefont {Szameit}}]{Weimann2016}%
  \BibitemOpen
  \bibfield  {author} {\bibinfo {author} {\bibfnamefont {S.}~\bibnamefont {Weimann}}, \bibinfo {author} {\bibfnamefont {M.}~\bibnamefont {Kremer}}, \bibinfo {author} {\bibfnamefont {Y.}~\bibnamefont {Plotnik}}, \bibinfo {author} {\bibfnamefont {Y.}~\bibnamefont {Lumer}}, \bibinfo {author} {\bibfnamefont {S.}~\bibnamefont {Nolte}}, \bibinfo {author} {\bibfnamefont {K.~G.}\ \bibnamefont {Makris}}, \bibinfo {author} {\bibfnamefont {M.}~\bibnamefont {Segev}}, \bibinfo {author} {\bibfnamefont {M. C.}\ \bibnamefont {Rechtsman}}, \ and\ \bibinfo {author} {\bibfnamefont {A.}~\bibnamefont {Szameit}},\ }\bibfield  {title} {\enquote {\bibinfo {title} {Topologically protected bound states in photonic parity{\textendash}time-symmetric crystals},}\ }\href {\doibase 10.1038/nmat4811} {\bibfield  {journal} {\bibinfo  {journal} {Nature Materials}\ }\textbf {\bibinfo {volume} {16}},\ \bibinfo {pages} {433--438} (\bibinfo {year} {2016})}\BibitemShut {NoStop}%
\bibitem [{\citenamefont {Vicencio}\ \emph {et~al.}(2024)\citenamefont {Vicencio}, \citenamefont {Rom\'an-Cort\'es}, \citenamefont {Rubio-Sald\'ias}, \citenamefont {Vildoso},\ and\ \citenamefont {Foa~Torres}}]{vicencioNR}%
  \BibitemOpen
  \bibfield  {author} {\bibinfo {author} {\bibfnamefont {R.A.}\ \bibnamefont {Vicencio}}, \bibinfo {author} {\bibfnamefont {D.}~\bibnamefont {Rom\'an-Cort\'es}}, \bibinfo {author} {\bibfnamefont {M.}~\bibnamefont {Rubio-Sald\'ias}}, \bibinfo {author} {\bibfnamefont {P.}~\bibnamefont {Vildoso}}, \ and\ \bibinfo {author} {\bibfnamefont {L.E.F.}\ \bibnamefont {Foa~Torres}},\ }\href {https://arxiv.org/abs/2407.18174} {\enquote {\bibinfo {title} {{Non-Reciprocal Coupling in Photonics}},}\ } (\bibinfo {year} {2024}),\ \Eprint {http://arxiv.org/abs/2407.18174} {arXiv:2407.18174 [physics.optics]} \BibitemShut {NoStop}%
\bibitem [{\citenamefont {Longhi}(2009)}]{Longhi2009Apr}%
  \BibitemOpen
  \bibfield  {author} {\bibinfo {author} {\bibfnamefont {S.}~\bibnamefont {Longhi}},\ }\bibfield  {title} {\enquote {\bibinfo {title} {{Quantum-optical analogies using photonic structures}},}\ }\href {\doibase 10.1002/lpor.200810055} {\bibfield  {journal} {\bibinfo  {journal} {Laser Photonics Rev.}\ }\textbf {\bibinfo {volume} {3}},\ \bibinfo {pages} {243--261} (\bibinfo {year} {2009})}\BibitemShut {NoStop}%
\bibitem [{\citenamefont {Mazanov}\ \emph {et~al.}(2024{\natexlab{a}})\citenamefont {Mazanov}, \citenamefont {Rom{\ifmmode\acute{a}\else\'{a}\fi}n-Cort{\ifmmode\acute{e}\else\'{e}\fi}s}, \citenamefont {C{\ifmmode\acute{a}\else\'{a}\fi}ceres-Aravena}, \citenamefont {Cid}, \citenamefont {Gorlach},\ and\ \citenamefont {Vicencio}}]{mol1us}%
  \BibitemOpen
  \bibfield  {author} {\bibinfo {author} {\bibfnamefont {Maxim}\ \bibnamefont {Mazanov}}, \bibinfo {author} {\bibfnamefont {Diego}\ \bibnamefont {Rom{\ifmmode\acute{a}\else\'{a}\fi}n-Cort{\ifmmode\acute{e}\else\'{e}\fi}s}}, \bibinfo {author} {\bibfnamefont {Gabriel}\ \bibnamefont {C{\ifmmode\acute{a}\else\'{a}\fi}ceres-Aravena}}, \bibinfo {author} {\bibfnamefont {Christofer}\ \bibnamefont {Cid}}, \bibinfo {author} {\bibfnamefont {Maxim~A.}\ \bibnamefont {Gorlach}}, \ and\ \bibinfo {author} {\bibfnamefont {Rodrigo~A.}\ \bibnamefont {Vicencio}},\ }\bibfield  {title} {\enquote {\bibinfo {title} {{Photonic Molecule Approach to Multiorbital Topology}},}\ }\href {\doibase 10.1021/acs.nanolett.4c00728} {\bibfield  {journal} {\bibinfo  {journal} {Nano Lett.}\ }\textbf {\bibinfo {volume} {24}},\ \bibinfo {pages} {4595--4601} (\bibinfo {year} {2024}{\natexlab{a}})}\BibitemShut {NoStop}%
\bibitem [{\citenamefont {Ozawa}\ \emph {et~al.}(2019)\citenamefont {Ozawa}, \citenamefont {Price}, \citenamefont {Amo}, \citenamefont {Goldman}, \citenamefont {Hafezi}, \citenamefont {Lu}, \citenamefont {Rechtsman}, \citenamefont {Schuster}, \citenamefont {Simon}, \citenamefont {Zilberberg},\ and\ \citenamefont {Carusotto}}]{Ozawa_RMP_2019}%
  \BibitemOpen
  \bibfield  {author} {\bibinfo {author} {\bibfnamefont {Tomoki}\ \bibnamefont {Ozawa}}, \bibinfo {author} {\bibfnamefont {Hannah~M.}\ \bibnamefont {Price}}, \bibinfo {author} {\bibfnamefont {Alberto}\ \bibnamefont {Amo}}, \bibinfo {author} {\bibfnamefont {Nathan}\ \bibnamefont {Goldman}}, \bibinfo {author} {\bibfnamefont {Mohammad}\ \bibnamefont {Hafezi}}, \bibinfo {author} {\bibfnamefont {Ling}\ \bibnamefont {Lu}}, \bibinfo {author} {\bibfnamefont {Mikael~C.}\ \bibnamefont {Rechtsman}}, \bibinfo {author} {\bibfnamefont {David}\ \bibnamefont {Schuster}}, \bibinfo {author} {\bibfnamefont {Jonathan}\ \bibnamefont {Simon}}, \bibinfo {author} {\bibfnamefont {Oded}\ \bibnamefont {Zilberberg}}, \ and\ \bibinfo {author} {\bibfnamefont {Iacopo}\ \bibnamefont {Carusotto}},\ }\bibfield  {title} {\enquote {\bibinfo {title} {Topological photonics},}\ }\href {\doibase 10.1103/RevModPhys.91.015006} {\bibfield  {journal} {\bibinfo  {journal} {Rev. Mod. Phys.}\ }\textbf {\bibinfo {volume} {91}},\ \bibinfo {pages} {015006}
  (\bibinfo {year} {2019})}\BibitemShut {NoStop}%
\bibitem [{\citenamefont {Khanikaev}\ \emph {et~al.}(2013)\citenamefont {Khanikaev}, \citenamefont {Hossein~Mousavi}, \citenamefont {Tse}, \citenamefont {Kargarian}, \citenamefont {MacDonald},\ and\ \citenamefont {Shvets}}]{Khanikaev2013Mar}%
  \BibitemOpen
  \bibfield  {author} {\bibinfo {author} {\bibfnamefont {Alexander~B.}\ \bibnamefont {Khanikaev}}, \bibinfo {author} {\bibfnamefont {S.}~\bibnamefont {Hossein~Mousavi}}, \bibinfo {author} {\bibfnamefont {Wang-Kong}\ \bibnamefont {Tse}}, \bibinfo {author} {\bibfnamefont {Mehdi}\ \bibnamefont {Kargarian}}, \bibinfo {author} {\bibfnamefont {Allan~H.}\ \bibnamefont {MacDonald}}, \ and\ \bibinfo {author} {\bibfnamefont {Gennady}\ \bibnamefont {Shvets}},\ }\bibfield  {title} {\enquote {\bibinfo {title} {{Photonic topological insulators}},}\ }\href {\doibase 10.1038/nmat3520} {\bibfield  {journal} {\bibinfo  {journal} {Nat. Mater.}\ }\textbf {\bibinfo {volume} {12}},\ \bibinfo {pages} {233--239} (\bibinfo {year} {2013})}\BibitemShut {NoStop}%
\bibitem [{\citenamefont {C\'aceres-Aravena}\ \emph {et~al.}(2020)\citenamefont {C\'aceres-Aravena}, \citenamefont {Torres},\ and\ \citenamefont {Vicencio}}]{PRASP1D}%
  \BibitemOpen
  \bibfield  {author} {\bibinfo {author} {\bibfnamefont {G.}~\bibnamefont {C\'aceres-Aravena}}, \bibinfo {author} {\bibfnamefont {L.~E. F.~Foa}\ \bibnamefont {Torres}}, \ and\ \bibinfo {author} {\bibfnamefont {R.~A.}\ \bibnamefont {Vicencio}},\ }\bibfield  {title} {\enquote {\bibinfo {title} {Topological and flat-band states induced by hybridized linear interactions in one-dimensional photonic lattices},}\ }\href {\doibase 10.1103/PhysRevA.102.023505} {\bibfield  {journal} {\bibinfo  {journal} {Phys. Rev. A}\ }\textbf {\bibinfo {volume} {102}},\ \bibinfo {pages} {023505} (\bibinfo {year} {2020})}\BibitemShut {NoStop}%
\bibitem [{\citenamefont {Savelev}\ and\ \citenamefont {Gorlach}(2020)}]{SavelevGorlach_PRB}%
  \BibitemOpen
  \bibfield  {author} {\bibinfo {author} {\bibfnamefont {Roman~S.}\ \bibnamefont {Savelev}}\ and\ \bibinfo {author} {\bibfnamefont {Maxim~A.}\ \bibnamefont {Gorlach}},\ }\bibfield  {title} {\enquote {\bibinfo {title} {Topological states in arrays of optical waveguides engineered via mode interference},}\ }\href {https://journals.aps.org/prb/abstract/10.1103/PhysRevB.102.161112} {\bibfield  {journal} {\bibinfo  {journal} {Physical Review B}\ }\textbf {\bibinfo {volume} {102}},\ \bibinfo {pages} {161112} (\bibinfo {year} {2020})}\BibitemShut {NoStop}%
\bibitem [{\citenamefont {Schulz}\ \emph {et~al.}(2022{\natexlab{a}})\citenamefont {Schulz}, \citenamefont {Noh}, \citenamefont {Benalcazar}, \citenamefont {Bahl},\ and\ \citenamefont {von Freymann}}]{Schulz2022Nov}%
  \BibitemOpen
  \bibfield  {author} {\bibinfo {author} {\bibfnamefont {Julian}\ \bibnamefont {Schulz}}, \bibinfo {author} {\bibfnamefont {Jiho}\ \bibnamefont {Noh}}, \bibinfo {author} {\bibfnamefont {Wladimir~A.}\ \bibnamefont {Benalcazar}}, \bibinfo {author} {\bibfnamefont {Gaurav}\ \bibnamefont {Bahl}}, \ and\ \bibinfo {author} {\bibfnamefont {Georg}\ \bibnamefont {von Freymann}},\ }\bibfield  {title} {\enquote {\bibinfo {title} {{Photonic quadrupole topological insulator using orbital-induced synthetic flux}},}\ }\href {\doibase 10.1038/s41467-022-33894-6} {\bibfield  {journal} {\bibinfo  {journal} {Nature Communications}\ }\textbf {\bibinfo {volume} {13}},\ \bibinfo {pages} {1--6} (\bibinfo {year} {2022}{\natexlab{a}})}\BibitemShut {NoStop}%
\bibitem [{\citenamefont {Mazanov}\ \emph {et~al.}(2024{\natexlab{b}})\citenamefont {Mazanov}, \citenamefont {Kupriianov}, \citenamefont {Savelev}, \citenamefont {He},\ and\ \citenamefont {Gorlach}}]{Mazanov2022}%
  \BibitemOpen
  \bibfield  {author} {\bibinfo {author} {\bibfnamefont {Maxim}\ \bibnamefont {Mazanov}}, \bibinfo {author} {\bibfnamefont {Anton~S.}\ \bibnamefont {Kupriianov}}, \bibinfo {author} {\bibfnamefont {Roman~S.}\ \bibnamefont {Savelev}}, \bibinfo {author} {\bibfnamefont {Zuxian}\ \bibnamefont {He}}, \ and\ \bibinfo {author} {\bibfnamefont {Maxim~A.}\ \bibnamefont {Gorlach}},\ }\bibfield  {title} {\enquote {\bibinfo {title} {Multipole higher-order topology in a multimode lattice},}\ }\href {\doibase 10.1103/physrevb.109.l201122} {\bibfield  {journal} {\bibinfo  {journal} {Physical Review B}\ }\textbf {\bibinfo {volume} {109}},\ \bibinfo {pages} {l201122} (\bibinfo {year} {2024}{\natexlab{b}})}\BibitemShut {NoStop}%
\bibitem [{\citenamefont {Cao}\ \emph {et~al.}(2018)\citenamefont {Cao}, \citenamefont {Fatemi}, \citenamefont {Fang}, \citenamefont {Watanabe}, \citenamefont {Taniguchi}, \citenamefont {Kaxiras},\ and\ \citenamefont {Jarillo-Herrero}}]{Cao2018}%
  \BibitemOpen
  \bibfield  {author} {\bibinfo {author} {\bibfnamefont {Yuan}\ \bibnamefont {Cao}}, \bibinfo {author} {\bibfnamefont {Valla}\ \bibnamefont {Fatemi}}, \bibinfo {author} {\bibfnamefont {Shiang}\ \bibnamefont {Fang}}, \bibinfo {author} {\bibfnamefont {Kenji}\ \bibnamefont {Watanabe}}, \bibinfo {author} {\bibfnamefont {Takashi}\ \bibnamefont {Taniguchi}}, \bibinfo {author} {\bibfnamefont {Efthimios}\ \bibnamefont {Kaxiras}}, \ and\ \bibinfo {author} {\bibfnamefont {Pablo}\ \bibnamefont {Jarillo-Herrero}},\ }\bibfield  {title} {\enquote {\bibinfo {title} {Unconventional superconductivity in magic-angle graphene superlattices},}\ }\href {\doibase 10.1038/nature26160} {\bibfield  {journal} {\bibinfo  {journal} {Nature}\ }\textbf {\bibinfo {volume} {556}},\ \bibinfo {pages} {43--50} (\bibinfo {year} {2018})}\BibitemShut {NoStop}%
\bibitem [{\citenamefont {Heikkil{\"a}}\ and\ \citenamefont {Volovik}(2016)}]{Heikki}%
  \BibitemOpen
  \bibfield  {author} {\bibinfo {author} {\bibfnamefont {Tero~T.}\ \bibnamefont {Heikkil{\"a}}}\ and\ \bibinfo {author} {\bibfnamefont {Grigory~E.}\ \bibnamefont {Volovik}},\ }\enquote {\bibinfo {title} {Flat bands as a route to high-temperature superconductivity in graphite},}\ in\ \href {\doibase 10.1007/978-3-319-39355-1_6} {\emph {\bibinfo {booktitle} {Basic Physics of Functionalized Graphite}}},\ \bibinfo {editor} {edited by\ \bibinfo {editor} {\bibfnamefont {Pablo~D.}\ \bibnamefont {Esquinazi}}}\ (\bibinfo  {publisher} {Springer International Publishing},\ \bibinfo {address} {Cham},\ \bibinfo {year} {2016})\ pp.\ \bibinfo {pages} {123--143}\BibitemShut {NoStop}%
\bibitem [{\citenamefont {Pyykk\"onen}\ \emph {et~al.}(2023)\citenamefont {Pyykk\"onen}, \citenamefont {Peotta},\ and\ \citenamefont {T\"orm\"a}}]{Torma1}%
  \BibitemOpen
  \bibfield  {author} {\bibinfo {author} {\bibfnamefont {Ville A.~J.}\ \bibnamefont {Pyykk\"onen}}, \bibinfo {author} {\bibfnamefont {Sebastiano}\ \bibnamefont {Peotta}}, \ and\ \bibinfo {author} {\bibfnamefont {P\"aivi}\ \bibnamefont {T\"orm\"a}},\ }\bibfield  {title} {\enquote {\bibinfo {title} {Suppression of nonequilibrium quasiparticle transport in flat-band superconductors},}\ }\href {\doibase 10.1103/PhysRevLett.130.216003} {\bibfield  {journal} {\bibinfo  {journal} {Phys. Rev. Lett.}\ }\textbf {\bibinfo {volume} {130}},\ \bibinfo {pages} {216003} (\bibinfo {year} {2023})}\BibitemShut {NoStop}%
\bibitem [{\citenamefont {Ciarrocchi}\ \emph {et~al.}(2022)\citenamefont {Ciarrocchi}, \citenamefont {Tagarelli}, \citenamefont {Avsar},\ and\ \citenamefont {Kis}}]{Ciarrocchi2022}%
  \BibitemOpen
  \bibfield  {author} {\bibinfo {author} {\bibfnamefont {Alberto}\ \bibnamefont {Ciarrocchi}}, \bibinfo {author} {\bibfnamefont {Fedele}\ \bibnamefont {Tagarelli}}, \bibinfo {author} {\bibfnamefont {Ahmet}\ \bibnamefont {Avsar}}, \ and\ \bibinfo {author} {\bibfnamefont {Andras}\ \bibnamefont {Kis}},\ }\bibfield  {title} {\enquote {\bibinfo {title} {{Excitonic devices with van der Waals heterostructures: valleytronics meets twistronics}},}\ }\href {\doibase 10.1038/s41578-021-00408-7} {\bibfield  {journal} {\bibinfo  {journal} {Nature Reviews Materials}\ }\textbf {\bibinfo {volume} {7}},\ \bibinfo {pages} {449--464} (\bibinfo {year} {2022})}\BibitemShut {NoStop}%
\bibitem [{\citenamefont {Hu}\ \emph {et~al.}(2021)\citenamefont {Hu}, \citenamefont {Qiu},\ and\ \citenamefont {Al{\`u}}}]{Hu2021}%
  \BibitemOpen
  \bibfield  {author} {\bibinfo {author} {\bibfnamefont {Guangwei}\ \bibnamefont {Hu}}, \bibinfo {author} {\bibfnamefont {Cheng-Wei}\ \bibnamefont {Qiu}}, \ and\ \bibinfo {author} {\bibfnamefont {Andrea}\ \bibnamefont {Al{\`u}}},\ }\bibfield  {title} {\enquote {\bibinfo {title} {Twistronics for photons: opinion},}\ }\href {\doibase 10.1364/ome.423521} {\bibfield  {journal} {\bibinfo  {journal} {Optical Materials Express}\ }\textbf {\bibinfo {volume} {11}},\ \bibinfo {pages} {1377} (\bibinfo {year} {2021})}\BibitemShut {NoStop}%
\bibitem [{\citenamefont {Zhou}\ \emph {et~al.}(2020)\citenamefont {Zhou}, \citenamefont {Lin}, \citenamefont {Lu}, \citenamefont {Lai}, \citenamefont {Hou},\ and\ \citenamefont {Jiang}}]{Hua_LPR_2020}%
  \BibitemOpen
  \bibfield  {author} {\bibinfo {author} {\bibfnamefont {Xiaoxi}\ \bibnamefont {Zhou}}, \bibinfo {author} {\bibfnamefont {Zhi-Kang}\ \bibnamefont {Lin}}, \bibinfo {author} {\bibfnamefont {Weixin}\ \bibnamefont {Lu}}, \bibinfo {author} {\bibfnamefont {Yun}\ \bibnamefont {Lai}}, \bibinfo {author} {\bibfnamefont {Bo}~\bibnamefont {Hou}}, \ and\ \bibinfo {author} {\bibfnamefont {Jian-Hua}\ \bibnamefont {Jiang}},\ }\bibfield  {title} {\enquote {\bibinfo {title} {{Twisted Quadrupole Topological Photonic Crystals}},}\ }\href {\doibase 10.1002/lpor.202000010} {\bibfield  {journal} {\bibinfo  {journal} {Laser Photonics Rev.}\ }\textbf {\bibinfo {volume} {14}},\ \bibinfo {pages} {2000010} (\bibinfo {year} {2020})}\BibitemShut {NoStop}%
\bibitem [{\citenamefont {Danieli}\ \emph {et~al.}(2024)\citenamefont {Danieli}, \citenamefont {Andreanov}, \citenamefont {Leykam},\ and\ \citenamefont {Flach}}]{finetuningFB}%
  \BibitemOpen
  \bibfield  {author} {\bibinfo {author} {\bibfnamefont {Carlo}\ \bibnamefont {Danieli}}, \bibinfo {author} {\bibfnamefont {Alexei}\ \bibnamefont {Andreanov}}, \bibinfo {author} {\bibfnamefont {Daniel}\ \bibnamefont {Leykam}}, \ and\ \bibinfo {author} {\bibfnamefont {Sergej}\ \bibnamefont {Flach}},\ }\bibfield  {title} {\enquote {\bibinfo {title} {Flat band fine-tuning and its photonic applications},}\ }\href {\doibase doi:10.1515/nanoph-2024-0135} {\bibfield  {journal} {\bibinfo  {journal} {Nanophotonics}\ }\textbf {\bibinfo {volume} {13}},\ \bibinfo {pages} {3925--3944} (\bibinfo {year} {2024})}\BibitemShut {NoStop}%
\bibitem [{\citenamefont {Danieli}\ \emph {et~al.}(2021)\citenamefont {Danieli}, \citenamefont {Andreanov}, \citenamefont {Mithun},\ and\ \citenamefont {Flach}}]{Danieli2021Aug1}%
  \BibitemOpen
  \bibfield  {author} {\bibinfo {author} {\bibfnamefont {Carlo}\ \bibnamefont {Danieli}}, \bibinfo {author} {\bibfnamefont {Alexei}\ \bibnamefont {Andreanov}}, \bibinfo {author} {\bibfnamefont {Thudiyangal}\ \bibnamefont {Mithun}}, \ and\ \bibinfo {author} {\bibfnamefont {Sergej}\ \bibnamefont {Flach}},\ }\bibfield  {title} {\enquote {\bibinfo {title} {{Nonlinear caging in all-bands-flat lattices}},}\ }\href {\doibase 10.1103/PhysRevB.104.085131} {\bibfield  {journal} {\bibinfo  {journal} {Phys. Rev. B}\ }\textbf {\bibinfo {volume} {104}},\ \bibinfo {pages} {085131} (\bibinfo {year} {2021})}\BibitemShut {NoStop}%
\bibitem [{\citenamefont {C\'aceres-Aravena}\ \emph {et~al.}(2024)\citenamefont {C\'aceres-Aravena}, \citenamefont {Nedi\ifmmode~\acute{c}\else \'{c}\fi{}}, \citenamefont {Vildoso}, \citenamefont {Gligori\ifmmode~\acute{c}\else \'{c}\fi{}}, \citenamefont {Petrovic}, \citenamefont {Maluckov},\ and\ \citenamefont {Vicencio}}]{CaceresPRL2024}%
  \BibitemOpen
  \bibfield  {author} {\bibinfo {author} {\bibfnamefont {Gabriel}\ \bibnamefont {C\'aceres-Aravena}}, \bibinfo {author} {\bibfnamefont {Milica}\ \bibnamefont {Nedi\ifmmode~\acute{c}\else \'{c}\fi{}}}, \bibinfo {author} {\bibfnamefont {Paloma}\ \bibnamefont {Vildoso}}, \bibinfo {author} {\bibfnamefont {Goran}\ \bibnamefont {Gligori\ifmmode~\acute{c}\else \'{c}\fi{}}}, \bibinfo {author} {\bibfnamefont {Jovana}\ \bibnamefont {Petrovic}}, \bibinfo {author} {\bibfnamefont {Aleksandra}\ \bibnamefont {Maluckov}}, \ and\ \bibinfo {author} {\bibfnamefont {Rodrigo~A.}\ \bibnamefont {Vicencio}},\ }\bibfield  {title} {\enquote {\bibinfo {title} {Compact topological edge states in flux-dressed graphenelike photonic lattices},}\ }\href {\doibase 10.1103/PhysRevLett.133.116304} {\bibfield  {journal} {\bibinfo  {journal} {Phys. Rev. Lett.}\ }\textbf {\bibinfo {volume} {133}},\ \bibinfo {pages} {116304} (\bibinfo {year} {2024})}\BibitemShut {NoStop}%
\bibitem [{\citenamefont {Flach}\ \emph {et~al.}(2014)\citenamefont {Flach}, \citenamefont {Leykam}, \citenamefont {Bodyfelt}, \citenamefont {Matthies},\ and\ \citenamefont {Desyatnikov}}]{Flach2014}%
  \BibitemOpen
  \bibfield  {author} {\bibinfo {author} {\bibfnamefont {Sergej}\ \bibnamefont {Flach}}, \bibinfo {author} {\bibfnamefont {Daniel}\ \bibnamefont {Leykam}}, \bibinfo {author} {\bibfnamefont {Joshua~D.}\ \bibnamefont {Bodyfelt}}, \bibinfo {author} {\bibfnamefont {Peter}\ \bibnamefont {Matthies}}, \ and\ \bibinfo {author} {\bibfnamefont {Anton~S.}\ \bibnamefont {Desyatnikov}},\ }\bibfield  {title} {\enquote {\bibinfo {title} {{Detangling flat bands into Fano lattices}},}\ }\href {\doibase 10.1209/0295-5075/105/30001} {\bibfield  {journal} {\bibinfo  {journal} {EPL (Europhysics Letters)}\ }\textbf {\bibinfo {volume} {105}},\ \bibinfo {pages} {30001} (\bibinfo {year} {2014})}\BibitemShut {NoStop}%
\bibitem [{\citenamefont {Maimaiti}\ \emph {et~al.}(2017)\citenamefont {Maimaiti}, \citenamefont {Andreanov}, \citenamefont {Park}, \citenamefont {Gendelman},\ and\ \citenamefont {Flach}}]{Maimaiti2017Mar}%
  \BibitemOpen
  \bibfield  {author} {\bibinfo {author} {\bibfnamefont {Wulayimu}\ \bibnamefont {Maimaiti}}, \bibinfo {author} {\bibfnamefont {Alexei}\ \bibnamefont {Andreanov}}, \bibinfo {author} {\bibfnamefont {Hee~Chul}\ \bibnamefont {Park}}, \bibinfo {author} {\bibfnamefont {Oleg}\ \bibnamefont {Gendelman}}, \ and\ \bibinfo {author} {\bibfnamefont {Sergej}\ \bibnamefont {Flach}},\ }\bibfield  {title} {\enquote {\bibinfo {title} {{Compact localized states and flat-band generators in one dimension}},}\ }\href {\doibase 10.1103/PhysRevB.95.115135} {\bibfield  {journal} {\bibinfo  {journal} {Phys. Rev. B}\ }\textbf {\bibinfo {volume} {95}},\ \bibinfo {pages} {115135} (\bibinfo {year} {2017})}\BibitemShut {NoStop}%
\bibitem [{\citenamefont {Maimaiti}\ \emph {et~al.}(2019)\citenamefont {Maimaiti}, \citenamefont {Flach},\ and\ \citenamefont {Andreanov}}]{Maimaiti2019Mar}%
  \BibitemOpen
  \bibfield  {author} {\bibinfo {author} {\bibfnamefont {Wulayimu}\ \bibnamefont {Maimaiti}}, \bibinfo {author} {\bibfnamefont {Sergej}\ \bibnamefont {Flach}}, \ and\ \bibinfo {author} {\bibfnamefont {Alexei}\ \bibnamefont {Andreanov}},\ }\bibfield  {title} {\enquote {\bibinfo {title} {{Universal $d=1$ flat band generator from compact localized states}},}\ }\href {\doibase 10.1103/PhysRevB.99.125129} {\bibfield  {journal} {\bibinfo  {journal} {Phys. Rev. B}\ }\textbf {\bibinfo {volume} {99}},\ \bibinfo {pages} {125129} (\bibinfo {year} {2019})}\BibitemShut {NoStop}%
\bibitem [{\citenamefont {Maimaiti}\ \emph {et~al.}(2021)\citenamefont {Maimaiti}, \citenamefont {Andreanov},\ and\ \citenamefont {Flach}}]{Maimaiti2021Apr}%
  \BibitemOpen
  \bibfield  {author} {\bibinfo {author} {\bibfnamefont {Wulayimu}\ \bibnamefont {Maimaiti}}, \bibinfo {author} {\bibfnamefont {Alexei}\ \bibnamefont {Andreanov}}, \ and\ \bibinfo {author} {\bibfnamefont {Sergej}\ \bibnamefont {Flach}},\ }\bibfield  {title} {\enquote {\bibinfo {title} {{Flat-band generator in two dimensions}},}\ }\href {\doibase 10.1103/PhysRevB.103.165116} {\bibfield  {journal} {\bibinfo  {journal} {Phys. Rev. B}\ }\textbf {\bibinfo {volume} {103}},\ \bibinfo {pages} {165116} (\bibinfo {year} {2021})}\BibitemShut {NoStop}%
\bibitem [{\citenamefont {Morales-Inostroza}\ and\ \citenamefont {Vicencio}(2016)}]{FBluis}%
  \BibitemOpen
  \bibfield  {author} {\bibinfo {author} {\bibfnamefont {Luis}\ \bibnamefont {Morales-Inostroza}}\ and\ \bibinfo {author} {\bibfnamefont {Rodrigo~A.}\ \bibnamefont {Vicencio}},\ }\bibfield  {title} {\enquote {\bibinfo {title} {Simple method to construct flat-band lattices},}\ }\href {\doibase 10.1103/PhysRevA.94.043831} {\bibfield  {journal} {\bibinfo  {journal} {Phys. Rev. A}\ }\textbf {\bibinfo {volume} {94}},\ \bibinfo {pages} {043831} (\bibinfo {year} {2016})}\BibitemShut {NoStop}%
\bibitem [{\citenamefont {Leykam}\ \emph {et~al.}(2018)\citenamefont {Leykam}, \citenamefont {Andreanov},\ and\ \citenamefont {Flach}}]{Leykam_APX}%
  \BibitemOpen
  \bibfield  {author} {\bibinfo {author} {\bibfnamefont {Daniel}\ \bibnamefont {Leykam}}, \bibinfo {author} {\bibfnamefont {Alexei}\ \bibnamefont {Andreanov}}, \ and\ \bibinfo {author} {\bibfnamefont {Sergej}\ \bibnamefont {Flach}},\ }\bibfield  {title} {\enquote {\bibinfo {title} {Artificial flat band systems: from lattice models to experiments},}\ }\href {\doibase https://doi.org/10.1080/23746149.2018.1473052} {\bibfield  {journal} {\bibinfo  {journal} {Advances in Physics: X}\ }\textbf {\bibinfo {volume} {3}},\ \bibinfo {pages} {1473052} (\bibinfo {year} {2018})}\BibitemShut {NoStop}%
\bibitem [{\citenamefont {Leykam}\ and\ \citenamefont {Flach}(2018)}]{Leykam2018Jul}%
  \BibitemOpen
  \bibfield  {author} {\bibinfo {author} {\bibfnamefont {Daniel}\ \bibnamefont {Leykam}}\ and\ \bibinfo {author} {\bibfnamefont {Sergej}\ \bibnamefont {Flach}},\ }\bibfield  {title} {\enquote {\bibinfo {title} {{Perspective: Photonic flatbands}},}\ }\href {\doibase 10.1063/1.5034365} {\bibfield  {journal} {\bibinfo  {journal} {APL Photonics}\ }\textbf {\bibinfo {volume} {3}},\ \bibinfo {pages} {070901} (\bibinfo {year} {2018})}\BibitemShut {NoStop}%
\bibitem [{\citenamefont {Vicencio~Poblete}(2021)}]{vicencioReportFB}%
  \BibitemOpen
  \bibfield  {author} {\bibinfo {author} {\bibfnamefont {Rodrigo~A.}\ \bibnamefont {Vicencio~Poblete}},\ }\bibfield  {title} {\enquote {\bibinfo {title} {Photonic flat band dynamics},}\ }\href {\doibase 10.1080/23746149.2021.1878057} {\bibfield  {journal} {\bibinfo  {journal} {Advances in Physics: X}\ }\textbf {\bibinfo {volume} {6}},\ \bibinfo {pages} {1878057} (\bibinfo {year} {2021})}\BibitemShut {NoStop}%
\bibitem [{\citenamefont {Zhu}\ \emph {et~al.}(2024)\citenamefont {Zhu}, \citenamefont {Zou}, \citenamefont {Ge}, \citenamefont {Wang}, \citenamefont {Cheng}, \citenamefont {Wang}, \citenamefont {Yuan}, \citenamefont {Sun}, \citenamefont {Xue},\ and\ \citenamefont {Zhang}}]{BZhang2024}%
  \BibitemOpen
  \bibfield  {author} {\bibinfo {author} {\bibfnamefont {Weiwei}\ \bibnamefont {Zhu}}, \bibinfo {author} {\bibfnamefont {Hongyu}\ \bibnamefont {Zou}}, \bibinfo {author} {\bibfnamefont {Yong}\ \bibnamefont {Ge}}, \bibinfo {author} {\bibfnamefont {Yin}\ \bibnamefont {Wang}}, \bibinfo {author} {\bibfnamefont {Zheyu}\ \bibnamefont {Cheng}}, \bibinfo {author} {\bibfnamefont {Bing-bing}\ \bibnamefont {Wang}}, \bibinfo {author} {\bibfnamefont {Shou-qi}\ \bibnamefont {Yuan}}, \bibinfo {author} {\bibfnamefont {Hong-xiang}\ \bibnamefont {Sun}}, \bibinfo {author} {\bibfnamefont {Haoran}\ \bibnamefont {Xue}}, \ and\ \bibinfo {author} {\bibfnamefont {Baile}\ \bibnamefont {Zhang}},\ }\bibfield  {title} {\enquote {\bibinfo {title} {{Flatbands from Bound States in the Continuum for Orbital Angular Momentum Localization}},}\ }\href {\doibase 10.48550/ARXIV.2410.04040} {\bibfield  {journal} {\bibinfo  {journal} {arXiv: 2410.04040}\ } (\bibinfo {year} {2024}),\ 10.48550/ARXIV.2410.04040},\ \Eprint
  {http://arxiv.org/abs/2410.04040} {arXiv:2410.04040 [physics.optics]} \BibitemShut {NoStop}%
\bibitem [{\citenamefont {J{\ifmmode\ddot{o}\else\"{o}\fi}rg}\ \emph {et~al.}(2020)\citenamefont {J{\ifmmode\ddot{o}\else\"{o}\fi}rg}, \citenamefont {Queralt{\ifmmode\acute{o}\else\'{o}\fi}}, \citenamefont {Kremer}, \citenamefont {Pelegr{\ifmmode\acute{\imath}\else\'{\i}\fi}}, \citenamefont {Schulz}, \citenamefont {Szameit}, \citenamefont {von Freymann}, \citenamefont {Mompart},\ and\ \citenamefont {Ahufinger}}]{Jorg2020Aug}%
  \BibitemOpen
  \bibfield  {author} {\bibinfo {author} {\bibfnamefont {Christina}\ \bibnamefont {J{\ifmmode\ddot{o}\else\"{o}\fi}rg}}, \bibinfo {author} {\bibfnamefont {Gerard}\ \bibnamefont {Queralt{\ifmmode\acute{o}\else\'{o}\fi}}}, \bibinfo {author} {\bibfnamefont {Mark}\ \bibnamefont {Kremer}}, \bibinfo {author} {\bibfnamefont {Gerard}\ \bibnamefont {Pelegr{\ifmmode\acute{\imath}\else\'{\i}\fi}}}, \bibinfo {author} {\bibfnamefont {Julian}\ \bibnamefont {Schulz}}, \bibinfo {author} {\bibfnamefont {Alexander}\ \bibnamefont {Szameit}}, \bibinfo {author} {\bibfnamefont {Georg}\ \bibnamefont {von Freymann}}, \bibinfo {author} {\bibfnamefont {Jordi}\ \bibnamefont {Mompart}}, \ and\ \bibinfo {author} {\bibfnamefont {Ver{\ifmmode\grave{o}\else\`{o}\fi}nica}\ \bibnamefont {Ahufinger}},\ }\bibfield  {title} {\enquote {\bibinfo {title} {{Artificial gauge field switching using orbital angular momentum modes in optical waveguides}},}\ }\href {\doibase 10.1038/s41377-020-00385-6} {\bibfield  {journal} {\bibinfo  {journal} {Light
  Sci. Appl.}\ }\textbf {\bibinfo {volume} {9}},\ \bibinfo {pages} {1--7} (\bibinfo {year} {2020})}\BibitemShut {NoStop}%
\bibitem [{\citenamefont {Pelegr{\ifmmode\acute{\imath}\else\'{\i}\fi}}\ \emph {et~al.}(2019)\citenamefont {Pelegr{\ifmmode\acute{\imath}\else\'{\i}\fi}}, \citenamefont {Marques}, \citenamefont {Dias}, \citenamefont {Daley}, \citenamefont {Mompart},\ and\ \citenamefont {Ahufinger}}]{Pelegri2019Feb}%
  \BibitemOpen
  \bibfield  {author} {\bibinfo {author} {\bibfnamefont {G.}~\bibnamefont {Pelegr{\ifmmode\acute{\imath}\else\'{\i}\fi}}}, \bibinfo {author} {\bibfnamefont {A.~M.}\ \bibnamefont {Marques}}, \bibinfo {author} {\bibfnamefont {R.~G.}\ \bibnamefont {Dias}}, \bibinfo {author} {\bibfnamefont {A.~J.}\ \bibnamefont {Daley}}, \bibinfo {author} {\bibfnamefont {J.}~\bibnamefont {Mompart}}, \ and\ \bibinfo {author} {\bibfnamefont {V.}~\bibnamefont {Ahufinger}},\ }\bibfield  {title} {\enquote {\bibinfo {title} {{Topological edge states and Aharanov-Bohm caging with ultracold atoms carrying orbital angular momentum}},}\ }\href {\doibase 10.1103/PhysRevA.99.023613} {\bibfield  {journal} {\bibinfo  {journal} {Phys. Rev. A}\ }\textbf {\bibinfo {volume} {99}},\ \bibinfo {pages} {023613} (\bibinfo {year} {2019})}\BibitemShut {NoStop}%
\bibitem [{\citenamefont {Mazanov}\ \emph {et~al.}(2023)\citenamefont {Mazanov}, \citenamefont {Kupriianov}, \citenamefont {He}, \citenamefont {Savelev},\ and\ \citenamefont {Gorlach}}]{MazanovIEEE}%
  \BibitemOpen
  \bibfield  {author} {\bibinfo {author} {\bibfnamefont {Maxim}\ \bibnamefont {Mazanov}}, \bibinfo {author} {\bibfnamefont {Anton}\ \bibnamefont {Kupriianov}}, \bibinfo {author} {\bibfnamefont {Zuxian}\ \bibnamefont {He}}, \bibinfo {author} {\bibfnamefont {Roman}\ \bibnamefont {Savelev}}, \ and\ \bibinfo {author} {\bibfnamefont {Maxim}\ \bibnamefont {Gorlach}},\ }\bibfield  {title} {\enquote {\bibinfo {title} {{Higher-Order Topology and Fully Flat Bands in Multimode Photonic Time-Reversal-Invariant Lattices}},}\ }in\ \href {\doibase 10.1109/IEEECONF59639.2023.10367218} {\emph {\bibinfo {booktitle} {{2023 Light Conference}}}}\ (\bibinfo  {publisher} {IEEE},\ \bibinfo {year} {2023})\ pp.\ \bibinfo {pages} {11--16}\BibitemShut {NoStop}%
\bibitem [{\citenamefont {Vidal}\ \emph {et~al.}(1998)\citenamefont {Vidal}, \citenamefont {Mosseri},\ and\ \citenamefont {Dou\ifmmode~\mbox{\c{c}}\else \c{c}\fi{}ot}}]{Vidal_PRL}%
  \BibitemOpen
  \bibfield  {author} {\bibinfo {author} {\bibfnamefont {Julien}\ \bibnamefont {Vidal}}, \bibinfo {author} {\bibfnamefont {R\'emy}\ \bibnamefont {Mosseri}}, \ and\ \bibinfo {author} {\bibfnamefont {Benoit}\ \bibnamefont {Dou\ifmmode~\mbox{\c{c}}\else \c{c}\fi{}ot}},\ }\bibfield  {title} {\enquote {\bibinfo {title} {{A}haronov-{B}ohm cages in two-dimensional structures},}\ }\href {\doibase 10.1103/PhysRevLett.81.5888} {\bibfield  {journal} {\bibinfo  {journal} {Phys. Rev. Lett.}\ }\textbf {\bibinfo {volume} {81}},\ \bibinfo {pages} {5888--5891} (\bibinfo {year} {1998})}\BibitemShut {NoStop}%
\bibitem [{\citenamefont {Longhi}(2014)}]{Longhi_OL}%
  \BibitemOpen
  \bibfield  {author} {\bibinfo {author} {\bibfnamefont {Stefano}\ \bibnamefont {Longhi}},\ }\bibfield  {title} {\enquote {\bibinfo {title} {{A}haronov-{B}ohm photonic cages in waveguide and coupled resonator lattices by synthetic magnetic fields},}\ }\href {\doibase https://doi.org/10.1364/OL.39.005892} {\bibfield  {journal} {\bibinfo  {journal} {Optics Letters}\ }\textbf {\bibinfo {volume} {39}},\ \bibinfo {pages} {5892--5895} (\bibinfo {year} {2014})}\BibitemShut {NoStop}%
\bibitem [{\citenamefont {Yang}\ \emph {et~al.}(2024)\citenamefont {Yang}, \citenamefont {Li}, \citenamefont {Yang}, \citenamefont {Xie}, \citenamefont {Zhang}, \citenamefont {Yuan}, \citenamefont {Cai}, \citenamefont {Wang},\ and\ \citenamefont {Gao}}]{Yang2024Feb}%
  \BibitemOpen
  \bibfield  {author} {\bibinfo {author} {\bibfnamefont {Jing}\ \bibnamefont {Yang}}, \bibinfo {author} {\bibfnamefont {Yuanzhen}\ \bibnamefont {Li}}, \bibinfo {author} {\bibfnamefont {Yumeng}\ \bibnamefont {Yang}}, \bibinfo {author} {\bibfnamefont {Xinrong}\ \bibnamefont {Xie}}, \bibinfo {author} {\bibfnamefont {Zijian}\ \bibnamefont {Zhang}}, \bibinfo {author} {\bibfnamefont {Jiale}\ \bibnamefont {Yuan}}, \bibinfo {author} {\bibfnamefont {Han}\ \bibnamefont {Cai}}, \bibinfo {author} {\bibfnamefont {Da-Wei}\ \bibnamefont {Wang}}, \ and\ \bibinfo {author} {\bibfnamefont {Fei}\ \bibnamefont {Gao}},\ }\bibfield  {title} {\enquote {\bibinfo {title} {{Realization of all-band-flat photonic lattices}},}\ }\href {\doibase 10.1038/s41467-024-45580-w} {\bibfield  {journal} {\bibinfo  {journal} {Nat. Commun.}\ }\textbf {\bibinfo {volume} {15}},\ \bibinfo {pages} {1--7} (\bibinfo {year} {2024})}\BibitemShut {NoStop}%
\bibitem [{\citenamefont {Hestand}\ and\ \citenamefont {Spano}(2018)}]{chemicalMA18}%
  \BibitemOpen
  \bibfield  {author} {\bibinfo {author} {\bibfnamefont {Nicholas~J.}\ \bibnamefont {Hestand}}\ and\ \bibinfo {author} {\bibfnamefont {Frank~C.}\ \bibnamefont {Spano}},\ }\bibfield  {title} {\enquote {\bibinfo {title} {Expanded theory of h- and j-molecular aggregates: The effects of vibronic coupling and intermolecular charge transfer},}\ }\href {\doibase 10.1021/acs.chemrev.7b00581} {\bibfield  {journal} {\bibinfo  {journal} {Chemical Reviews}\ }\textbf {\bibinfo {volume} {118}},\ \bibinfo {pages} {7069--7163} (\bibinfo {year} {2018})},\ \bibinfo {note} {pMID: 29664617},\ \Eprint {http://arxiv.org/abs/https://doi.org/10.1021/acs.chemrev.7b00581} {https://doi.org/10.1021/acs.chemrev.7b00581} \BibitemShut {NoStop}%
\bibitem [{\citenamefont {Kazmaier}\ and\ \citenamefont {Hoffmann}(1994)}]{Kazmaier1994Oct}%
  \BibitemOpen
  \bibfield  {author} {\bibinfo {author} {\bibfnamefont {Peter~M.}\ \bibnamefont {Kazmaier}}\ and\ \bibinfo {author} {\bibfnamefont {Roald}\ \bibnamefont {Hoffmann}},\ }\bibfield  {title} {\enquote {\bibinfo {title} {{A Theoretical Study of Crystallochromy. Quantum Interference Effects in the Spectra of Perylene Pigments}},}\ }\href {\doibase 10.1021/ja00100a038} {\bibfield  {journal} {\bibinfo  {journal} {J. Am. Chem. Soc.}\ }\textbf {\bibinfo {volume} {116}},\ \bibinfo {pages} {9684--9691} (\bibinfo {year} {1994})}\BibitemShut {NoStop}%
\bibitem [{\citenamefont {Yariv}\ and\ \citenamefont {Yeh}(2003)}]{yariv_optical_2003}%
  \BibitemOpen
  \bibfield  {author} {\bibinfo {author} {\bibfnamefont {Amnon}\ \bibnamefont {Yariv}}\ and\ \bibinfo {author} {\bibfnamefont {Pochi}\ \bibnamefont {Yeh}},\ }\href@noop {} {\emph {\bibinfo {title} {Optical waves in crystals: propagation and control of laser radiation}}},\ \bibinfo {edition} {wiley classics library ed}\ ed.,\ Wiley classics library\ (\bibinfo  {publisher} {John Wiley and Sons},\ \bibinfo {address} {Hoboken, N.J},\ \bibinfo {year} {2003})\BibitemShut {NoStop}%
\bibitem [{Sup()}]{Supplement}%
  \BibitemOpen
  \href@noop {} {\enquote {\bibinfo {title} {{See Supplemental Material at [url will be inserted by publisher] for the numerical calculation of couplings from the eigenmodes of a dimer; evaluation of non-orthogonality corrections for waveguide lattices; construction of Bloch Hamiltonian and Wannier center calculation; computation of the inverse participation ratio; description of the femtosecond laser writing technique; further details on experiments.}}}\ }\BibitemShut {NoStop}%
\bibitem [{\citenamefont {Maczewsky}\ \emph {et~al.}(2019)\citenamefont {Maczewsky}, \citenamefont {Weimann}, \citenamefont {Kremer}, \citenamefont {Heinrich},\ and\ \citenamefont {Szameit}}]{Maczewsky}%
  \BibitemOpen
  \bibfield  {author} {\bibinfo {author} {\bibfnamefont {Lukas~J.}\ \bibnamefont {Maczewsky}}, \bibinfo {author} {\bibfnamefont {Steffen}\ \bibnamefont {Weimann}}, \bibinfo {author} {\bibfnamefont {Mark}\ \bibnamefont {Kremer}}, \bibinfo {author} {\bibfnamefont {Matthias}\ \bibnamefont {Heinrich}}, \ and\ \bibinfo {author} {\bibfnamefont {Alexander}\ \bibnamefont {Szameit}},\ }\bibfield  {title} {\enquote {\bibinfo {title} {{Experimental study of non-orthogonal modes in tight-binding lattices}},}\ }in\ \href {\doibase 10.1364/CLEO_QELS.2019.FM1C.6} {\emph {\bibinfo {booktitle} {{2019 Conference on Lasers and Electro-Optics (CLEO)}}}}\ (\bibinfo  {publisher} {IEEE},\ \bibinfo {year} {2019})\ pp.\ \bibinfo {pages} {05--10}\BibitemShut {NoStop}%
\bibitem [{\citenamefont {Schulz}\ \emph {et~al.}(2022{\natexlab{b}})\citenamefont {Schulz}, \citenamefont {J{\ifmmode\ddot{o}\else\"{o}\fi}rg}, \citenamefont {J{\ifmmode\ddot{o}\else\"{o}\fi}rg}, \citenamefont {von Freymann},\ and\ \citenamefont {von Freymann}}]{Schulz2022Mar}%
  \BibitemOpen
  \bibfield  {author} {\bibinfo {author} {\bibfnamefont {J.}~\bibnamefont {Schulz}}, \bibinfo {author} {\bibfnamefont {C.}~\bibnamefont {J{\ifmmode\ddot{o}\else\"{o}\fi}rg}}, \bibinfo {author} {\bibfnamefont {C.}~\bibnamefont {J{\ifmmode\ddot{o}\else\"{o}\fi}rg}}, \bibinfo {author} {\bibfnamefont {G.}~\bibnamefont {von Freymann}}, \ and\ \bibinfo {author} {\bibfnamefont {G.}~\bibnamefont {von Freymann}},\ }\bibfield  {title} {\enquote {\bibinfo {title} {{Geometric control of next-nearest-neighbor coupling in evanescently coupled dielectric waveguides}},}\ }\href {\doibase 10.1364/OE.447921} {\bibfield  {journal} {\bibinfo  {journal} {Opt. Express}\ }\textbf {\bibinfo {volume} {30}},\ \bibinfo {pages} {9869--9877} (\bibinfo {year} {2022}{\natexlab{b}})}\BibitemShut {NoStop}%
\bibitem [{\citenamefont {Thouless}(1974)}]{Thouless_1974}%
  \BibitemOpen
  \bibfield  {author} {\bibinfo {author} {\bibfnamefont {D.J.}\ \bibnamefont {Thouless}},\ }\bibfield  {title} {\enquote {\bibinfo {title} {Electrons in disordered systems and the theory of localization},}\ }\href {\doibase https://doi.org/10.1016/0370-1573(74)90029-5} {\bibfield  {journal} {\bibinfo  {journal} {Physics Reports}\ }\textbf {\bibinfo {volume} {13}},\ \bibinfo {pages} {93--142} (\bibinfo {year} {1974})}\BibitemShut {NoStop}%
\bibitem [{\citenamefont {Leykam}\ \emph {et~al.}(2012)\citenamefont {Leykam}, \citenamefont {Bahat-Treidel},\ and\ \citenamefont {Desyatnikov}}]{LiebLeykam}%
  \BibitemOpen
  \bibfield  {author} {\bibinfo {author} {\bibfnamefont {Daniel}\ \bibnamefont {Leykam}}, \bibinfo {author} {\bibfnamefont {Omri}\ \bibnamefont {Bahat-Treidel}}, \ and\ \bibinfo {author} {\bibfnamefont {Anton~S.}\ \bibnamefont {Desyatnikov}},\ }\bibfield  {title} {\enquote {\bibinfo {title} {{Pseudospin and nonlinear conical diffraction in Lieb lattices}},}\ }\href {\doibase 10.1103/PhysRevA.86.031805} {\bibfield  {journal} {\bibinfo  {journal} {Phys. Rev. A}\ }\textbf {\bibinfo {volume} {86}},\ \bibinfo {pages} {031805} (\bibinfo {year} {2012})}\BibitemShut {NoStop}%
\bibitem [{\citenamefont {Vicencio}\ \emph {et~al.}(2015)\citenamefont {Vicencio}, \citenamefont {Cantillano}, \citenamefont {Morales-Inostroza}, \citenamefont {Real}, \citenamefont {Mej\'{\i}a-Cort\'es}, \citenamefont {Weimann}, \citenamefont {Szameit},\ and\ \citenamefont {Molina}}]{Liebus}%
  \BibitemOpen
  \bibfield  {author} {\bibinfo {author} {\bibfnamefont {Rodrigo~A.}\ \bibnamefont {Vicencio}}, \bibinfo {author} {\bibfnamefont {Camilo}\ \bibnamefont {Cantillano}}, \bibinfo {author} {\bibfnamefont {Luis}\ \bibnamefont {Morales-Inostroza}}, \bibinfo {author} {\bibfnamefont {Basti\'an}\ \bibnamefont {Real}}, \bibinfo {author} {\bibfnamefont {Cristian}\ \bibnamefont {Mej\'{\i}a-Cort\'es}}, \bibinfo {author} {\bibfnamefont {Steffen}\ \bibnamefont {Weimann}}, \bibinfo {author} {\bibfnamefont {Alexander}\ \bibnamefont {Szameit}}, \ and\ \bibinfo {author} {\bibfnamefont {Mario~I.}\ \bibnamefont {Molina}},\ }\bibfield  {title} {\enquote {\bibinfo {title} {{Observation of Localized States in Lieb Photonic Lattices}},}\ }\href {\doibase 10.1103/PhysRevLett.114.245503} {\bibfield  {journal} {\bibinfo  {journal} {Phys. Rev. Lett.}\ }\textbf {\bibinfo {volume} {114}},\ \bibinfo {pages} {245503} (\bibinfo {year} {2015})}\BibitemShut {NoStop}%
\bibitem [{\citenamefont {Mukherjee}\ \emph {et~al.}(2015)\citenamefont {Mukherjee}, \citenamefont {Spracklen}, \citenamefont {Choudhury}, \citenamefont {Goldman}, \citenamefont {\"Ohberg}, \citenamefont {Andersson},\ and\ \citenamefont {Thomson}}]{LiebSeba}%
  \BibitemOpen
  \bibfield  {author} {\bibinfo {author} {\bibfnamefont {Sebabrata}\ \bibnamefont {Mukherjee}}, \bibinfo {author} {\bibfnamefont {Alexander}\ \bibnamefont {Spracklen}}, \bibinfo {author} {\bibfnamefont {Debaditya}\ \bibnamefont {Choudhury}}, \bibinfo {author} {\bibfnamefont {Nathan}\ \bibnamefont {Goldman}}, \bibinfo {author} {\bibfnamefont {Patrik}\ \bibnamefont {\"Ohberg}}, \bibinfo {author} {\bibfnamefont {Erika}\ \bibnamefont {Andersson}}, \ and\ \bibinfo {author} {\bibfnamefont {Robert~R.}\ \bibnamefont {Thomson}},\ }\bibfield  {title} {\enquote {\bibinfo {title} {{Observation of a Localized Flat-Band State in a Photonic Lieb Lattice}},}\ }\href {\doibase 10.1103/PhysRevLett.114.245504} {\bibfield  {journal} {\bibinfo  {journal} {Phys. Rev. Lett.}\ }\textbf {\bibinfo {volume} {114}},\ \bibinfo {pages} {245504} (\bibinfo {year} {2015})}\BibitemShut {NoStop}%
\bibitem [{\citenamefont {Mukherjee}\ and\ \citenamefont {Thomson}(2017)}]{SebaOL}%
  \BibitemOpen
  \bibfield  {author} {\bibinfo {author} {\bibfnamefont {Sebabrata}\ \bibnamefont {Mukherjee}}\ and\ \bibinfo {author} {\bibfnamefont {Robert~R.}\ \bibnamefont {Thomson}},\ }\bibfield  {title} {\enquote {\bibinfo {title} {Observation of robust flat-band localization in driven photonic rhombic lattices},}\ }\href {\doibase 10.1364/OL.42.002243} {\bibfield  {journal} {\bibinfo  {journal} {Opt. Lett.}\ }\textbf {\bibinfo {volume} {42}},\ \bibinfo {pages} {2243--2246} (\bibinfo {year} {2017})}\BibitemShut {NoStop}%
\bibitem [{\citenamefont {Su}\ \emph {et~al.}(1979)\citenamefont {Su}, \citenamefont {Schrieffer},\ and\ \citenamefont {Heeger}}]{1DSSH}%
  \BibitemOpen
  \bibfield  {author} {\bibinfo {author} {\bibfnamefont {W.~P.}\ \bibnamefont {Su}}, \bibinfo {author} {\bibfnamefont {J.~R.}\ \bibnamefont {Schrieffer}}, \ and\ \bibinfo {author} {\bibfnamefont {A.~J.}\ \bibnamefont {Heeger}},\ }\bibfield  {title} {\enquote {\bibinfo {title} {Solitons in polyacetylene},}\ }\href {https://journals.aps.org/prl/abstract/10.1103/PhysRevLett.42.1698} {\bibfield  {journal} {\bibinfo  {journal} {Phys. Rev. Lett.}\ }\textbf {\bibinfo {volume} {42}},\ \bibinfo {pages} {1698--1701} (\bibinfo {year} {1979})}\BibitemShut {NoStop}%
\bibitem [{\citenamefont {Benalcazar}\ \emph {et~al.}(2019)\citenamefont {Benalcazar}, \citenamefont {Li},\ and\ \citenamefont {Hughes}}]{Quantization_2019}%
  \BibitemOpen
  \bibfield  {author} {\bibinfo {author} {\bibfnamefont {Wladimir~A.}\ \bibnamefont {Benalcazar}}, \bibinfo {author} {\bibfnamefont {Tianhe}\ \bibnamefont {Li}}, \ and\ \bibinfo {author} {\bibfnamefont {Taylor~L.}\ \bibnamefont {Hughes}},\ }\bibfield  {title} {\enquote {\bibinfo {title} {Quantization of fractional corner charge in ${C}_{n}$-symmetric higher-order topological crystalline insulators},}\ }\href {https://journals.aps.org/prb/abstract/10.1103/PhysRevB.99.245151} {\bibfield  {journal} {\bibinfo  {journal} {Phys. Rev. B}\ }\textbf {\bibinfo {volume} {99}},\ \bibinfo {pages} {245151} (\bibinfo {year} {2019})}\BibitemShut {NoStop}%
\end{thebibliography}%


\begin{thebibliography}{8}%
\makeatletter
\providecommand \@ifxundefined [1]{%
 \@ifx{#1\undefined}
}%
\providecommand \@ifnum [1]{%
 \ifnum #1\expandafter \@firstoftwo
 \else \expandafter \@secondoftwo
 \fi
}%
\providecommand \@ifx [1]{%
 \ifx #1\expandafter \@firstoftwo
 \else \expandafter \@secondoftwo
 \fi
}%
\providecommand \natexlab [1]{#1}%
\providecommand \enquote  [1]{``#1''}%
\providecommand \bibnamefont  [1]{#1}%
\providecommand \bibfnamefont [1]{#1}%
\providecommand \citenamefont [1]{#1}%
\providecommand \href@noop [0]{\@secondoftwo}%
\providecommand \href [0]{\begingroup \@sanitize@url \@href}%
\providecommand \@href[1]{\@@startlink{#1}\@@href}%
\providecommand \@@href[1]{\endgroup#1\@@endlink}%
\providecommand \@sanitize@url [0]{\catcode `\\12\catcode `\$12\catcode `\&12\catcode `\#12\catcode `\^12\catcode `\_12\catcode `\%12\relax}%
\providecommand \@@startlink[1]{}%
\providecommand \@@endlink[0]{}%
\providecommand \url  [0]{\begingroup\@sanitize@url \@url }%
\providecommand \@url [1]{\endgroup\@href {#1}{\urlprefix }}%
\providecommand \urlprefix  [0]{URL }%
\providecommand \Eprint [0]{\href }%
\providecommand \doibase [0]{https://doi.org/}%
\providecommand \selectlanguage [0]{\@gobble}%
\providecommand \bibinfo  [0]{\@secondoftwo}%
\providecommand \bibfield  [0]{\@secondoftwo}%
\providecommand \translation [1]{[#1]}%
\providecommand \BibitemOpen [0]{}%
\providecommand \bibitemStop [0]{}%
\providecommand \bibitemNoStop [0]{.\EOS\space}%
\providecommand \EOS [0]{\spacefactor3000\relax}%
\providecommand \BibitemShut  [1]{\csname bibitem#1\endcsname}%
\let\auto@bib@innerbib\@empty
\bibitem [{\citenamefont {Benalcazar}\ \emph {et~al.}(2017{\natexlab{a}})\citenamefont {Benalcazar}, \citenamefont {Bernevig},\ and\ \citenamefont {Hughes}}]{Benalcazar_2017_PRB}%
  \BibitemOpen
  \bibfield  {author} {\bibinfo {author} {\bibfnamefont {W.~A.}\ \bibnamefont {Benalcazar}}, \bibinfo {author} {\bibfnamefont {B.~A.}\ \bibnamefont {Bernevig}},\ and\ \bibinfo {author} {\bibfnamefont {T.~L.}\ \bibnamefont {Hughes}},\ }\bibfield  {title} {\bibinfo {title} {Electric multipole moments, topological multipole moment pumping, and chiral hinge states in crystalline insulators},\ }\href {https://journals.aps.org/prb/abstract/10.1103/PhysRevB.96.245115} {\bibfield  {journal} {\bibinfo  {journal} {Phys. Rev. B}\ }\textbf {\bibinfo {volume} {96}},\ \bibinfo {pages} {245115} (\bibinfo {year} {2017}{\natexlab{a}})}\BibitemShut {NoStop}%
\bibitem [{\citenamefont {Benalcazar}\ \emph {et~al.}(2017{\natexlab{b}})\citenamefont {Benalcazar}, \citenamefont {Bernevig},\ and\ \citenamefont {Hughes}}]{Benalcazar_2017_Science}%
  \BibitemOpen
  \bibfield  {author} {\bibinfo {author} {\bibfnamefont {W.~A.}\ \bibnamefont {Benalcazar}}, \bibinfo {author} {\bibfnamefont {B.~A.}\ \bibnamefont {Bernevig}},\ and\ \bibinfo {author} {\bibfnamefont {T.~L.}\ \bibnamefont {Hughes}},\ }\bibfield  {title} {\bibinfo {title} {Quantized electric multipole insulators},\ }\href {https://www.science.org/doi/abs/10.1126/science.aah6442} {\bibfield  {journal} {\bibinfo  {journal} {Science}\ }\textbf {\bibinfo {volume} {357}},\ \bibinfo {pages} {61} (\bibinfo {year} {2017}{\natexlab{b}})}\BibitemShut {NoStop}%
\bibitem [{\citenamefont {Davis}\ \emph {et~al.}(1996)\citenamefont {Davis}, \citenamefont {Miura}, \citenamefont {Sugimoto},\ and\ \citenamefont {Hirao}}]{Davis96}%
  \BibitemOpen
  \bibfield  {author} {\bibinfo {author} {\bibfnamefont {K.~M.}\ \bibnamefont {Davis}}, \bibinfo {author} {\bibfnamefont {K.}~\bibnamefont {Miura}}, \bibinfo {author} {\bibfnamefont {N.}~\bibnamefont {Sugimoto}},\ and\ \bibinfo {author} {\bibfnamefont {K.}~\bibnamefont {Hirao}},\ }\bibfield  {title} {\bibinfo {title} {Writing waveguides in glass with a femtosecond laser},\ }\href {https://doi.org/10.1364/OL.21.001729} {\bibfield  {journal} {\bibinfo  {journal} {Opt. Lett.}\ }\textbf {\bibinfo {volume} {21}},\ \bibinfo {pages} {1729} (\bibinfo {year} {1996})}\BibitemShut {NoStop}%
\bibitem [{\citenamefont {Szameit}\ \emph {et~al.}(2005)\citenamefont {Szameit}, \citenamefont {Bl\"{o}mer}, \citenamefont {Burghoff}, \citenamefont {Schreiber}, \citenamefont {Pertsch}, \citenamefont {Nolte}, \citenamefont {T\"{u}nnermann},\ and\ \citenamefont {Lederer}}]{Szameit2005}%
  \BibitemOpen
  \bibfield  {author} {\bibinfo {author} {\bibfnamefont {A.}~\bibnamefont {Szameit}}, \bibinfo {author} {\bibfnamefont {D.}~\bibnamefont {Bl\"{o}mer}}, \bibinfo {author} {\bibfnamefont {J.}~\bibnamefont {Burghoff}}, \bibinfo {author} {\bibfnamefont {T.}~\bibnamefont {Schreiber}}, \bibinfo {author} {\bibfnamefont {T.}~\bibnamefont {Pertsch}}, \bibinfo {author} {\bibfnamefont {S.}~\bibnamefont {Nolte}}, \bibinfo {author} {\bibfnamefont {A.}~\bibnamefont {T\"{u}nnermann}},\ and\ \bibinfo {author} {\bibfnamefont {F.}~\bibnamefont {Lederer}},\ }\bibfield  {title} {\bibinfo {title} {Discrete nonlinear localization in femtosecond laser written waveguides in fused silica},\ }\href {https://doi.org/10.1364/OPEX.13.010552} {\bibfield  {journal} {\bibinfo  {journal} {Opt. Express}\ }\textbf {\bibinfo {volume} {13}},\ \bibinfo {pages} {10552} (\bibinfo {year} {2005})}\BibitemShut {NoStop}%
\bibitem [{\citenamefont {Flamini}\ \emph {et~al.}(2015)\citenamefont {Flamini}, \citenamefont {Magrini}, \citenamefont {Rab}, \citenamefont {Spagnolo}, \citenamefont {D{\textquotesingle}Ambrosio}, \citenamefont {Mataloni}, \citenamefont {Sciarrino}, \citenamefont {Zandrini}, \citenamefont {Crespi}, \citenamefont {Ramponi},\ and\ \citenamefont {Osellame}}]{Flamini2015}%
  \BibitemOpen
  \bibfield  {author} {\bibinfo {author} {\bibfnamefont {F.}~\bibnamefont {Flamini}}, \bibinfo {author} {\bibfnamefont {L.}~\bibnamefont {Magrini}}, \bibinfo {author} {\bibfnamefont {A.~S.}\ \bibnamefont {Rab}}, \bibinfo {author} {\bibfnamefont {N.}~\bibnamefont {Spagnolo}}, \bibinfo {author} {\bibfnamefont {V.}~\bibnamefont {D{\textquotesingle}Ambrosio}}, \bibinfo {author} {\bibfnamefont {P.}~\bibnamefont {Mataloni}}, \bibinfo {author} {\bibfnamefont {F.}~\bibnamefont {Sciarrino}}, \bibinfo {author} {\bibfnamefont {T.}~\bibnamefont {Zandrini}}, \bibinfo {author} {\bibfnamefont {A.}~\bibnamefont {Crespi}}, \bibinfo {author} {\bibfnamefont {R.}~\bibnamefont {Ramponi}},\ and\ \bibinfo {author} {\bibfnamefont {R.}~\bibnamefont {Osellame}},\ }\bibfield  {title} {\bibinfo {title} {Thermally reconfigurable quantum photonic circuits at telecom wavelength by femtosecond laser micromachining},\ }\href {https://doi.org/10.1038/lsa.2015.127} {\bibfield  {journal} {\bibinfo  {journal} {Light: Science {\&} Applications}\
  }\textbf {\bibinfo {volume} {4}},\ \bibinfo {pages} {e354} (\bibinfo {year} {2015})}\BibitemShut {NoStop}%
\bibitem [{\citenamefont {Guzm\'an-Silva}\ \emph {et~al.}(2021)\citenamefont {Guzm\'an-Silva}, \citenamefont {C\'aceres-Aravena},\ and\ \citenamefont {Vicencio}}]{SPprl}%
  \BibitemOpen
  \bibfield  {author} {\bibinfo {author} {\bibfnamefont {D.}~\bibnamefont {Guzm\'an-Silva}}, \bibinfo {author} {\bibfnamefont {G.}~\bibnamefont {C\'aceres-Aravena}},\ and\ \bibinfo {author} {\bibfnamefont {R.~A.}\ \bibnamefont {Vicencio}},\ }\bibfield  {title} {\bibinfo {title} {Experimental observation of interorbital coupling},\ }\href {https://doi.org/10.1103/PhysRevLett.127.066601} {\bibfield  {journal} {\bibinfo  {journal} {Phys. Rev. Lett.}\ }\textbf {\bibinfo {volume} {127}},\ \bibinfo {pages} {066601} (\bibinfo {year} {2021})}\BibitemShut {NoStop}%
\bibitem [{\citenamefont {Vicencio}\ \emph {et~al.}(2015)\citenamefont {Vicencio}, \citenamefont {Cantillano}, \citenamefont {Morales-Inostroza}, \citenamefont {Real}, \citenamefont {Mej\'{\i}a-Cort\'es}, \citenamefont {Weimann}, \citenamefont {Szameit},\ and\ \citenamefont {Molina}}]{Liebus}%
  \BibitemOpen
  \bibfield  {author} {\bibinfo {author} {\bibfnamefont {R.~A.}\ \bibnamefont {Vicencio}}, \bibinfo {author} {\bibfnamefont {C.}~\bibnamefont {Cantillano}}, \bibinfo {author} {\bibfnamefont {L.}~\bibnamefont {Morales-Inostroza}}, \bibinfo {author} {\bibfnamefont {B.}~\bibnamefont {Real}}, \bibinfo {author} {\bibfnamefont {C.}~\bibnamefont {Mej\'{\i}a-Cort\'es}}, \bibinfo {author} {\bibfnamefont {S.}~\bibnamefont {Weimann}}, \bibinfo {author} {\bibfnamefont {A.}~\bibnamefont {Szameit}},\ and\ \bibinfo {author} {\bibfnamefont {M.~I.}\ \bibnamefont {Molina}},\ }\bibfield  {title} {\bibinfo {title} {{Observation of Localized States in Lieb Photonic Lattices}},\ }\href {https://doi.org/10.1103/PhysRevLett.114.245503} {\bibfield  {journal} {\bibinfo  {journal} {Phys. Rev. Lett.}\ }\textbf {\bibinfo {volume} {114}},\ \bibinfo {pages} {245503} (\bibinfo {year} {2015})}\BibitemShut {NoStop}%
\bibitem [{\citenamefont {Cáceres-Aravena}\ \emph {et~al.}(2023)\citenamefont {Cáceres-Aravena}, \citenamefont {Real}, \citenamefont {Guzmán-Silva}, \citenamefont {Vildoso}, \citenamefont {Salinas}, \citenamefont {Amo}, \citenamefont {Ozawa},\ and\ \citenamefont {Vicencio}}]{SpectralTransfer}%
  \BibitemOpen
  \bibfield  {author} {\bibinfo {author} {\bibfnamefont {G.}~\bibnamefont {Cáceres-Aravena}}, \bibinfo {author} {\bibfnamefont {B.}~\bibnamefont {Real}}, \bibinfo {author} {\bibfnamefont {D.}~\bibnamefont {Guzmán-Silva}}, \bibinfo {author} {\bibfnamefont {P.}~\bibnamefont {Vildoso}}, \bibinfo {author} {\bibfnamefont {I.}~\bibnamefont {Salinas}}, \bibinfo {author} {\bibfnamefont {A.}~\bibnamefont {Amo}}, \bibinfo {author} {\bibfnamefont {T.}~\bibnamefont {Ozawa}},\ and\ \bibinfo {author} {\bibfnamefont {R.~A.}\ \bibnamefont {Vicencio}},\ }\bibfield  {title} {\bibinfo {title} {{Edge-to-edge topological spectral transfer in diamond photonic lattices}},\ }\href {https://doi.org/10.1063/5.0153770} {\bibfield  {journal} {\bibinfo  {journal} {APL Photonics}\ }\textbf {\bibinfo {volume} {8}},\ \bibinfo {pages} {080801} (\bibinfo {year} {2023})}\BibitemShut {NoStop}%
\end{thebibliography}%





\end{document}


\title{Supplemental Materials:\\ Observation of the magic angle and flat band physics in dipolar photonic lattices}

\author{Diego Rom\'an-Cort\'es}
\email{These authors have equally contributed to this work}
\affiliation{Departamento de Física and Millenium Institute for Research in Optics–MIRO, Facultad de Ciencias Físicas y Matemáticas, Universidad de Chile, 8370448 Santiago, Chile}

\author{Maxim Mazanov}
\email{These authors have equally contributed to this work}
\affiliation{School of Physics and Engineering, ITMO University, Saint  Petersburg 197101, Russia}

\author{Rodrigo A. Vicencio}
\email{rvicencio@uchile.cl}
\affiliation{Departamento de Física and Millenium Institute for Research in Optics–MIRO, Facultad de Ciencias Físicas y Matemáticas, Universidad de Chile, 8370448 Santiago, Chile}

\author{Maxim A. Gorlach}
\email{m.gorlach@metalab.ifmo.ru}
\affiliation{School of Physics and Engineering, ITMO University, Saint  Petersburg 197101, Russia}

\maketitle

\onecolumngrid

\setcounter{equation}{0}
\setcounter{figure}{0}
\setcounter{table}{0}
\setcounter{page}{1}
\setcounter{section}{0}
\makeatletter
\renewcommand{\theequation}{S\arabic{equation}}
\renewcommand{\thefigure}{S\arabic{figure}}
\renewcommand{\bibnumfmt}[1]{[S#1]}
\renewcommand{\citenumfont}[1]{S#1}

\tableofcontents

\section{I. Numerical calculation of couplings from dimer eigenmodes}

\begin{figure*}[ht!]
	\centering
	\includegraphics[width=0.99\textwidth]{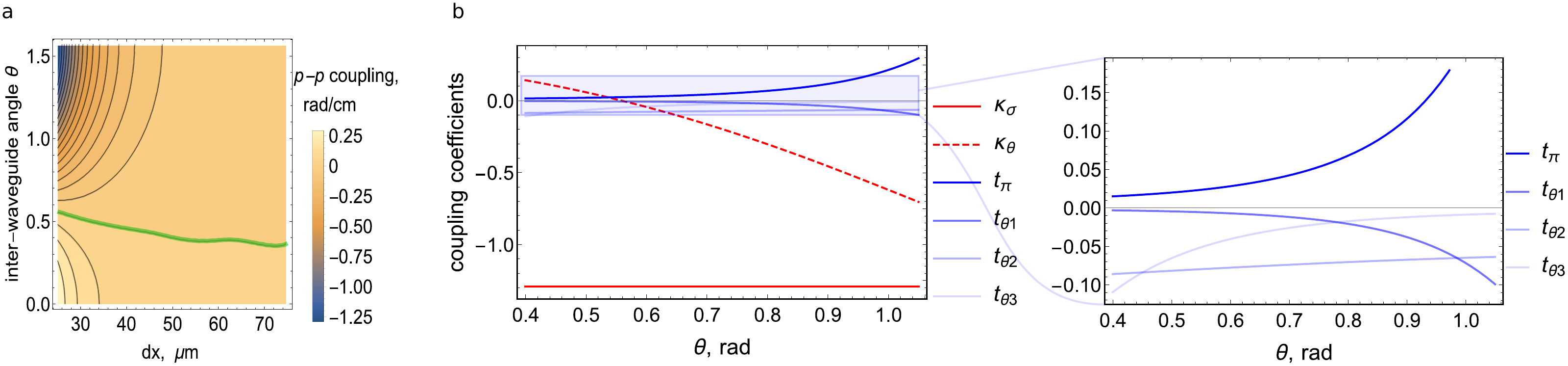}
	\caption{ 
        \textbf{a} Inter-waveguide coupling as a function of angle and distance. 
        \textbf{b} Couplings relevant for the strained honeycomb lattice. 
        }
        \label{fig:SF1}
\end{figure*}

We first numerically extract the $p$-mode coupling constant $\kappa(dx, \theta, \lambda)$ from $p$-mode waveguide dimer eigenmodes $k_0 \pm \kappa(dx, \theta, \lambda)$ in COMSOL Multiphysics for varying inter-waveguide distances $dx$, wavelengths $\lambda$ and the inter-waveguide angle $\theta$. 
The results for varying distances and angles are shown in Fig.~\ref{fig:SF1}~\textbf{a}, where the magic angle (zero coupling) line is highlighted in green. 
Next, we calculate the couplings relevant to our model which follow from the concrete geometry of the strained honeycomb lattice with fixed closest inter-waveguide distance $d$ and angle $\theta$: 
nearest-neighbour 
couplings $\kappa_\theta$ and $\kappa_\sigma$, two next-nearest-neighbour 
couplings $t_\pi$ and $t_{\theta2}$, and two next-next-nearest-neighbour (further-range) 
couplings $t_{\theta1}$ and $t_{\theta3}$ (for the fixed operating wavelength) 
\begin{eqnarray}
\label{couplings_lattice}
    && \kappa_\sigma = \kappa\left( d , \pi/2 \right)
    , \nonumber \\ 
    && \kappa_\theta = \kappa\left( d , \theta \right)
    , \nonumber \\ 
    && t_\pi = \kappa\left( 2 d \cos\theta , 0 \right)
    , \nonumber \\ 
    && t_{\theta1} = \kappa\left( d \sqrt{4 \cos^2 \theta + 1 }, \arctan\left(\frac{1}{2 \cos \theta}\right) \right)
    , \nonumber \\ 
    && t_{\theta2} = \kappa\left( d \sqrt{2 (\sin\theta + 1)}, \arctan\left(\frac{\sin \theta + 1}{\cos \theta}\right) \right)
    , \nonumber \\ 
    && t_{\theta3} = \kappa\left( d (2 \sin\theta + 1), \frac{\pi}{2} \right)
. 
\end{eqnarray}
The results are shown in Fig.~\ref{fig:SF1}~\textbf{b}. Note that although the long-range couplings are one order of magnitude smaller than the NN couplings, they become important and are responsible for the nonzero band dispersion near the magic angle.

\section{II. Numerical calculation of overlap integrals for the non-orthogonality corrections}

\begin{figure*}[ht!]
	\centering
	\includegraphics[width=0.99\textwidth]{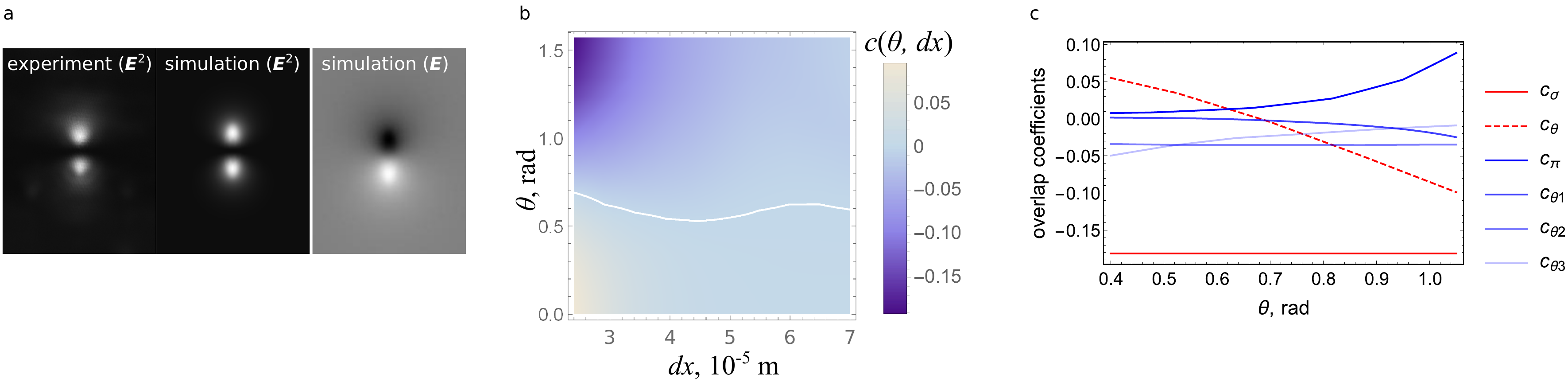}
	\caption{ 
        \textbf{a} Experimentally measured and numerically simulated intensity profiles (left, center), and simulated electric field  amplitude profiles (right). 
        \textbf{b} Overlap integrals as functions of inter-waveguide distance $dx$ and angle $\theta$ for the operating wavelength $730$~nm. 
        \textbf{c} Overlap integrals relevant for the strained honeycomb lattice. 
        }
        \label{fig:SF2}
\end{figure*}

The overlap integrals $\hat{c}_{\text{n.o.}}^{i, j}$ (Eq.(2) in the main text) relevant to the lattice configuration could be calculated similarly to Eqs.~\eqref{couplings_lattice}, while $\kappa$ is substituted by the distance- and angle-varying overlap function (for the fixed wavelength) for the waveguide dimers, calculated numerically in Wolfram Mathematica from the isolated waveguide field profile from COMSOL Multiphysics shown in Fig.~\ref{fig:SF2}\textbf{a} (note that numerical waveguide parameters ensured the reasonable matching of experimental and numerical intensity profiles; small deviations are due to the step-wise nature of the numerical waveguide refractive index profile). The results are shown in Fig.~\ref{fig:SF2}\textbf{b}-\textbf{c}. Interestingly, the overlap function also features a zero overlap region which physically corresponds to the situation when non-orthogonality corrections vanish. However, the conditions for the zero overlap do not coincide with the magic angle condition for the couplings. 
The calculated overlap integrals are then inserted into the matrix $\hat{c}_{\text{n.o.}}^{-1}$ appearing in the evolution equation Eq.~(3) in the main text; note that $\hat{c}_{\text{n.o.}}$ has the same structure as the full Hamiltonian $\hat{H}_\Sigma$. 


\section{III. The Bloch Hamiltonian and Wannier center calculation}

\begin{figure*}[ht!]
	\centering
	\includegraphics[width=0.85\textwidth]{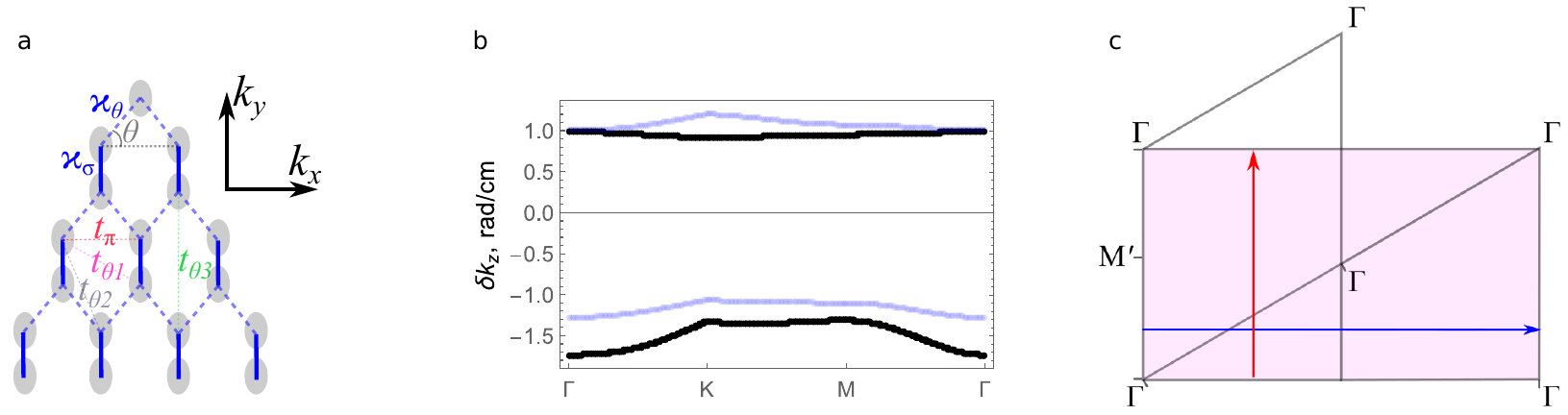}
	\caption{
        \textbf{a} Geometry of the lattice with couplings included in the full Hamiltonian~\eqref{fullH}. 
	    \textbf{b} Bulk bands of the full Bloch Hamiltonian $\hat{H}_\Sigma$            r the magic angle ($\theta_m = 0.56$~rad) and operating wavelength      $730$~nm: without the non-orthogonal correction (light blue) and with     the non-orthogonal correction matrix $\hat{c}_{n.o.}^{-1}$ (black).  
            \textbf{c} Wilson line directions (depicted as red and blue lines) and double-periodic enlarged Brilloin zone (highlighted in red) for our $C_2$-symmetric model, with a choice of periodic Wilson lines associated with position operators $\hat{x},\hat{y}$. 
	}
	\label{SfigWLoops}
\end{figure*}

The full Hamiltonian, which includes nearest-neighbour (NN), next-nearest-neighbour (NNN) and next-next-nearest-neighbour (NNNN) couplings follows from the model shown in Fig.2\textbf{e} in the main text, and reads: 
\begin{eqnarray}
\label{fullH}
    \hat{H}_\Sigma &=& 
    \left(
    \begin{array}{cc}
    0 & e^{-i \left(k_x + k_y \sqrt{3}\right)/2} \kappa_\sigma + ( 1 + e^{-i k_x})\kappa_\theta \\
    e^{i \left(k_x + k_y \sqrt{3}\right)/2} \kappa_\sigma + ( 1 + e^{i k_x})\kappa_\theta & 0 \\
\end{array}
\right)
+
    t_\pi \left(
    \begin{array}{cc}
    2 \cos{k_x} & 0 \\
    0 & 2 \cos{k_x} \\
    \end{array}
    \right)
+ \nonumber\\
&+&
    t_{\theta 1} \left(
    \begin{array}{cc}
     0 & e^{- i \left(3 k_x + k_y \sqrt{3}\right)/2} \left(1 + e^{2 i k_x}\right) \\
     e^{ i \left(3 k_x + k_y \sqrt{3}\right)/2} \left(1 + e^{- 2 i k_x}\right) & 0 \\
    \end{array}
    \right)
+ 
    t_{\theta 2} \left(
    \begin{array}{cc}
     4 \cos{\frac{k_x}{2}} \cos{\frac{\sqrt{3} k_y}{2}} & 0 \\
     0 & 4 \cos{\frac{k_x}{2}} \cos{\frac{\sqrt{3} k_y}{2}} \\
    \end{array}
    \right)
+ \nonumber\\
&+&
    t_{\theta 3} \left(
    \begin{array}{cc}
     0 & e^{- i \left(k_x - \sqrt{3} k_y \right)/2} \\
     e^{i \left(k_x - \sqrt{3} k_y \right)/2} & 0 \\
    \end{array}
    \right)
, 
\end{eqnarray}
where $\textbf{k} = \{k_x, k_y\}$ is the Bloch wavevector, see Fig.~\ref{SfigWLoops}\textbf{a}. An example of bulk Bloch spectrum for the periodic boundary conditions which follow from such Hamiltonian near the magic angle condition is depicted in Fig.~\ref{SfigWLoops}\textbf{b}. 

Using this Hamiltonian, we next calculate the Wannier centers by the Wilson loop technique~\cite{Benalcazar_2017_PRB,Benalcazar_2017_Science}. 
We choose a rectangular enlarged double-periodic region in the Brilloin zone and calculate the Wilson loops along the blue and red lines in Fig.~\ref{SfigWLoops}\textbf{c} to obtain Wannier center positions (bulk polarization projections) along $x$ and $y$, respectively. This calculation (including for the realistic experimental couplings) gives flat Wannier bands and a value of two-dimensional bulk polarization $\textbf{P} = \{ 0.5, 0.5 \}$ quantized by the inversion symmetry in the bulk. Additionally, while the value of $P_y = 0.5$ responsible for the appearance of the topological corner state in Fig.4 in the main text could be inferred independently from limiting uncoupled-1D-SSH argument depicted in Fig.4a, a similar argument can be made for $P_x = 0.5$ by performing an alternative decomposition of the lattice into tilted 1D SSH stripes. However, the nonzero value of $P_x$ manifests itself only in the even-layer cut of the lattice as a corner state, where it hybridizes with the edge states connected to the bulk polarization component $P_y$, and is not observed in our experiments when the lattice includes odd number of rows.

\section{IV. IPR with consecutive Hamiltonian approximations}

\begin{figure*}[ht!]
	\centering
	\includegraphics[width=0.35\textwidth]{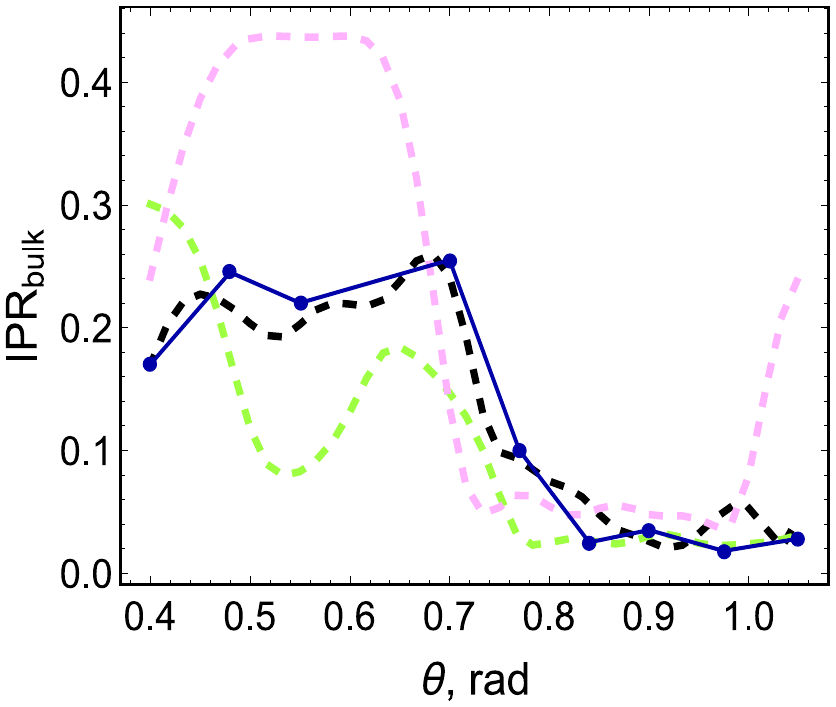}
	\caption{
		IPR for the bulk excitation for the experiment (dots with a line as a guide to the eye) and simulated numerically in three approximations: without the long-range couplings and non-orthogonal corrections (pink dashed curve), with the long-range couplings and without non-orthogonal corrections (green dashed curve), and with both the long-range couplings and non-orthogonal corrections (black dashed curve). 
	}
	\label{SFigUncorrectedCouplings}
\end{figure*}

Fig.~\ref{SFigUncorrectedCouplings} shows three IPR curves for the bulk excitation compared to the experimental data. 
This comparison highlights that both the long-range couplings and non-orthogonal corrections (black dashed curve) are needed to correctly reproduce the experimental results. 

\section{V. Femtosecond laser writing technique}

Along this work, we fabricate different photonic structures by using the femtosecond (fs) laser writing technique~\cite{Davis96,Szameit2005,Flamini2015}, as it is sketched in Fig.2(a) of the main text. Ultrashort pulses from an ATSEVA ANTAUS Yb-doped $1030$ nm fiber laser, at a repetition rate of $500$ kHz, are tightly focused on a borosilicate glass wafer (with refractive index $n_0 = 1.48$). The laser pulses weakly modify the material properties at the focusing region, inducing a small and permanent refractive index contrast of $\Delta n\approx 10^{-4}-10^{-3}$~\cite{SPprl}. Straight waveguides, along the glass wafer, are created by translating the glass along the $z$ propagation coordinate by means of a motorized XYZ Aerotech stage. Writing power and writing velocity are the critical parameters in this technique, such that the waveguide structure could support one or more states at a given excitation wavelength. In our samples, we used a writing velocity of $0.8$ mm/s and a nominal setup writing power of $\sim 14$ mW for the specific goal of having well formed $p$ modes, as the ones described in the main text, over a broad wavelength range around $730$ nm. 

\begin{figure*}[h]
	\centering
	\includegraphics[width=0.99\textwidth]{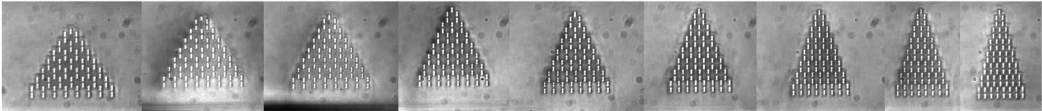}
	\caption{ 
        Microscopic black and white images of the input face of strained graphene-like photonic lattices. The aspect ratio of the lattice changes because of the angle sweep.
        }
        \label{fig:SFB&W}
\end{figure*}

After $p$-mode coupling characterization (described below and in the main text), a set of $p$-modes strained graphene-like lattices, of 71 waveguides each, were fabricated on a 70 mm long borosilicate glass wafer. Microscope images of them, after white light illumination, are shown in Fig.~\ref{fig:SFB&W}. 

\section{VI. Characterization setups}

In this work, we use two characterization setups, which are based in two very different experimental techniques, as we describe below:

\begin{figure*}[h]
	\centering
	\includegraphics[width=0.6\textwidth]{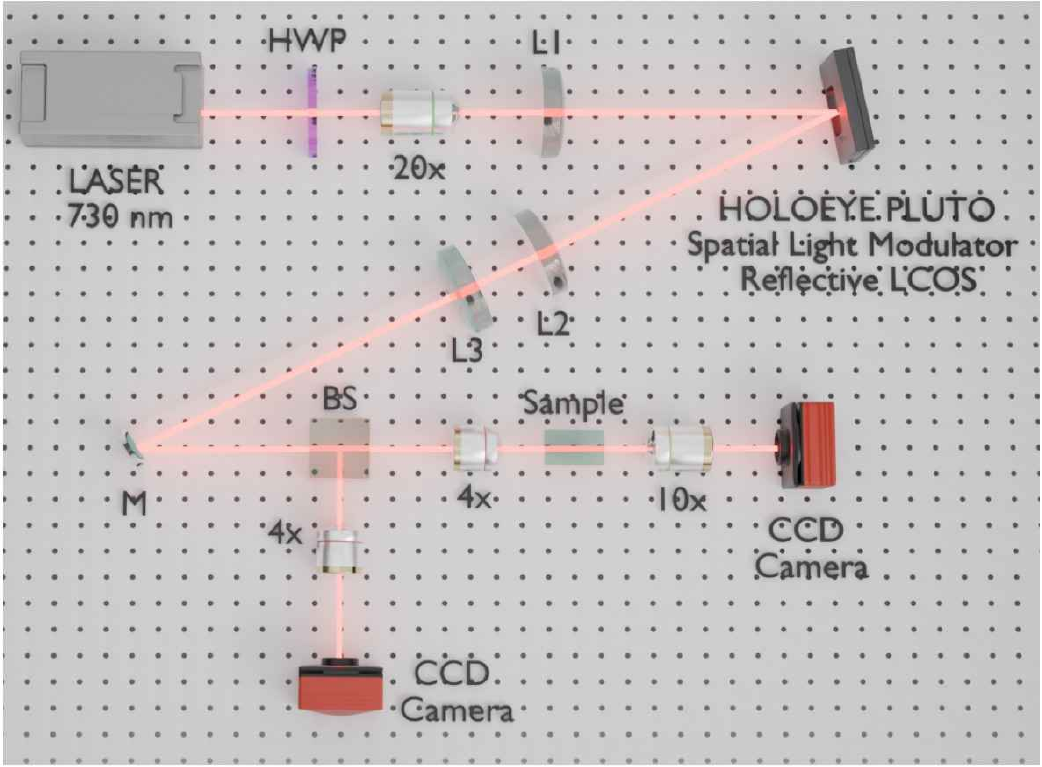}
	\caption{ 
        Spatial Light Modulation setup. HWP: Polarizer and Half-wave plate. L1: plano-convex lens focus 150 mm, L2: plano-convex lens focus 1000mm, L3: plano-convex lens focus 100 mm, M: Mirror, BS: 50/50 Beam Splitter.}
        \label{fig:SFSLMsetup}
\end{figure*}

\subsection{SLM-technique.} A Spatial Light Modulation (SLM) setup, as depicted in Fig.~\ref{fig:SFSLMsetup}, was implemented for a pure P-mode excitation. A simultaneous amplitude and phase modulation of a wide optical beam allows us to excite a given photonic structure with a specific and complex input image~\cite{Liebus}. This setup can be divided into four parts:

\begin{enumerate}
    \item Premodulation: A half-wave plate (HWP) after a polarizer (not shown) were used to match the polarization response of the SLM display. Collimated expansion of the laser beam was achieved using a 20x microscope objective and a 150 mm focal length plano-convex lens (L1).
    \item Modulation: A blazed grating, optimized for a wavelength of 730 nm, was used. An amplitude mask and a phase mask were then applied simultaneously to the beam.
    \item Input image condition: The modulated image was reduced in size $10$ times by means of a Keplerian telescope of focal lengths 1000 mm (L2) and 100 mm (L3). The image was then focused through a 4x microscope objective onto the input face of the glass wafer. The angles of the sample relative to the modulated input image were calibrated, so that the latter could be reflected back at the glass input facet and pass through the beam splitter (BS), perpendicular to the original direction. The image was remagnified by another 4x objective lens and captured by a CCD camera to observe and control the input conditions. The input face of the sample was illuminated using white light in order to match the size of the input image (a dipole profile in this case) to the one of the waveguide.
    \item Output Recording: The resulting propagation through the system is finally magnified by a 10x microscope objetive and captured on a CCD camera beam profiler.

\subsection{Supercontinuum (SC) laser excitation}

A wavelength-scan method~\cite{SpectralTransfer} was used to characterize the dynamical response of the lattice for a broader wavelength range, considering the dependence of the magic angle with respect to the input laser wavelength. As sketched in Fig.~\ref{fig:SFSCsetup}, an YSL SC-5 supercontinuum laser source connected to an acoustic modulator AOTF-PRO is used to select wavelengths in the interval $\{700,760\}$ nm. The beam is then polarized by means of a polarizer and a HWP to select horizontal (parallel to tabletop) polarization, and finally focused into the glass input facet by means of a 10x microscope objective. A fine tuning of the glass sample XYZ position as well as of the input beam focus is performed, such that we effectively excite the $p$-mode of a selected lattice multi-mode waveguide. Although only Gaussian-like profiles can be excited with this method, the $p$-mode band is far enough from the $s$-mode band such that mostly $p$-states are observed in the experiments.

\begin{figure*}[h]
	\centering
	\includegraphics[width=0.7\textwidth]{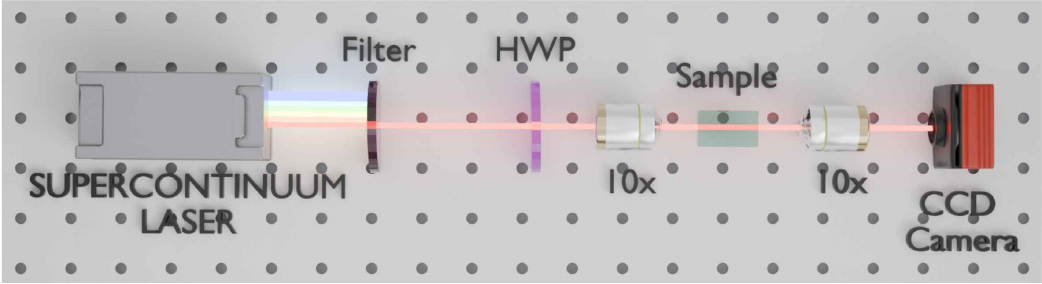}
	\caption{ 
       Supercontinuum setup. Filter: Acoustic modulator and iris. HWP: Half-wave plate.           }
\label{fig:SFSCsetup}
\end{figure*}


\section{VII. Full coupler results}

A set of 21 $p$-mode couplers, having a separation distance of 25 $\mu$m, were fabricated on a $L=30$ mm long glass wafer. Each coupler configuration consists of a full length excitation waveguide plus a shorter waveguide with a propagation distance, in this case, of 25 mm. At this separation distance the coupling constants are, in general, small, and this coupler configuration allows us to efficiently extract the respective values before a first oscillation cycle. The coupler angle $\theta$ was defined relative to the horizontal axis from 0 to $\pi/2$ radians, with a step of $\sim 0.08$ rad. All the output intensity images used for the experimental coupling data extraction, some of them displayed also in Fig.2c of the main text, are shown in Fig.~\ref{fig:SFdimerangle}.

\end{enumerate}
\begin{figure*}[h]
	\centering
	\includegraphics[width=0.9\textwidth]{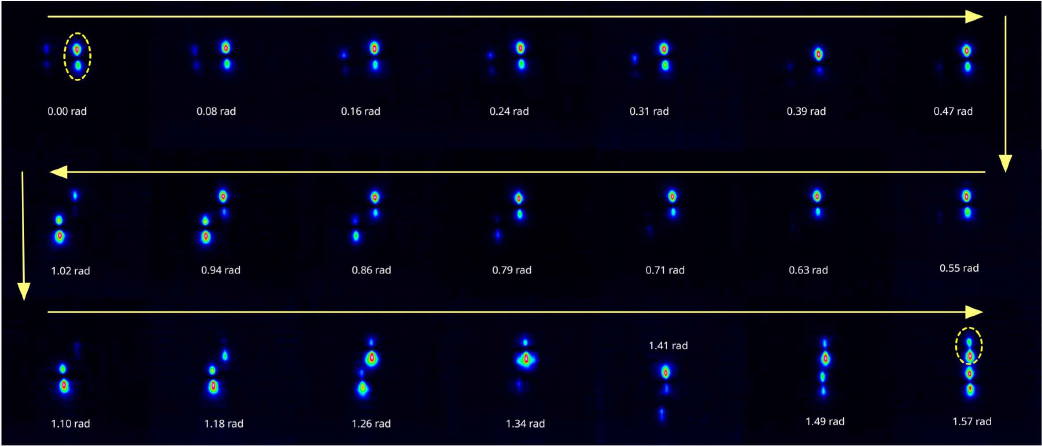}
	\caption{ 
        Coupler angle sweep from 0 to $\pi/2$ radians. The excitation $p$-mode waveguide corresponds to the top-right site (see yellow ellipses). 
        }
        \label{fig:SFdimerangle}
\end{figure*}

We observe a very clear and smooth evolution of the intensity at the excited (right-hand) waveguide with respect to the shorter (left) waveguide, over the coupler angle $\theta$. First of all, the amount of transferred energy at the non-excited waveguide slowly decreases up to a minimum at $\theta\sim0.55$ rad. Then, a faster increment of the intensity at this waveguide is observed with a perfect transfer around $\theta\sim1.10$ rad, for which the coupling constant is $\kappa\sim\pi/5$ cm$^{-1}$. Finally, the light starts to return to the input waveguide and a $\sim 50/50$ beam splitter is achieved for $\theta\sim\pi/2$ rad.

\section{VIII. SLM excitation of strained graphene-like lattices}

\begin{figure*}[h]
	\centering
	\includegraphics[width=0.9\textwidth]{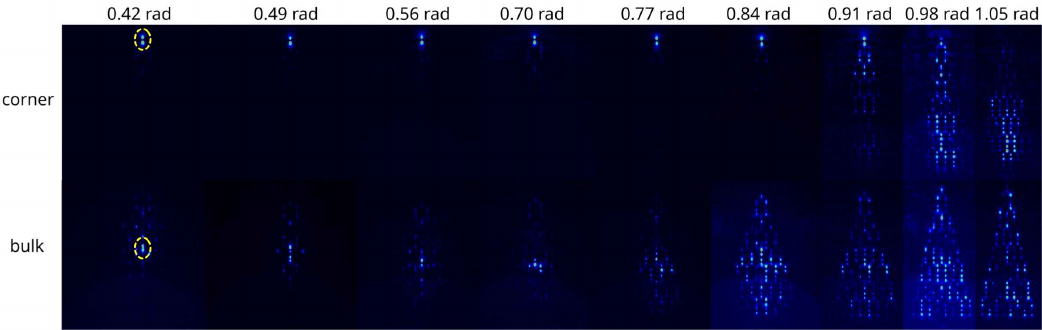}
	\caption{ 
        SLM dipole input condition for corner and bulk sites (see yellow ellipses). 
        }
        \label{fig:SFSLMexcitations}
\end{figure*}

Single-site $p$-mode excitation is achieved by means of the SLM technique at an excitation wavelength of $730$ nm. Specifically, we excited the upper corner and the bulk center for each of the nine lattices, as shown in Fig.~\ref{fig:SFSLMexcitations}. For a corner-site excitation, there is a very clear prevalence of corner localization up to an angle $\theta\sim0.84$ rad, with output profiles which look perfectly localized at the excitation region. For $\theta=0.91$ rad, we observe the emergence of a weak tail plus some background radiation, as an indication of the transition of this corner topological state and its connection with the upper band [see Fig.3a of the main text]. For $\theta>0.91$ rad, we do not observe corner localization but diffraction through the system.

On the other hand, a bulk central excitation diffracts only weakly for smaller angles and it becomes very well trapped at around the magic angle, where the waveguide pairs (dimers) decoupled from the rest of the lattice. Then, for an increasing angle $\theta$, all the couplings constants grow and next-nearest neighbour couplings also become relevant in the dynamics, allowing the light to diffract across the lattice. In this last case, we observe a clear phase transition into the magic angle dynamical regime, with the expected 2D caging effect due to an ABF regime.


\section{IX. Bulk and Corner SC excitation of strained graphene-like lattices}

Single-site excitation of our lattices by means of a SC laser source allows us to study the performance of the system over a larger set of parameters. In this case, as the $p$-mode coupling depends on the specific wavelength, including the magic angle regime, we can not use this scan method as the dynamical effective propagation~\cite{SpectralTransfer}. However, it nevertheless shows us the robustness of our observation depending on excitation wavelength, which can be useful for all-optical applications.

\begin{figure*}[h]
	\centering
	\includegraphics[width=0.49\textwidth]{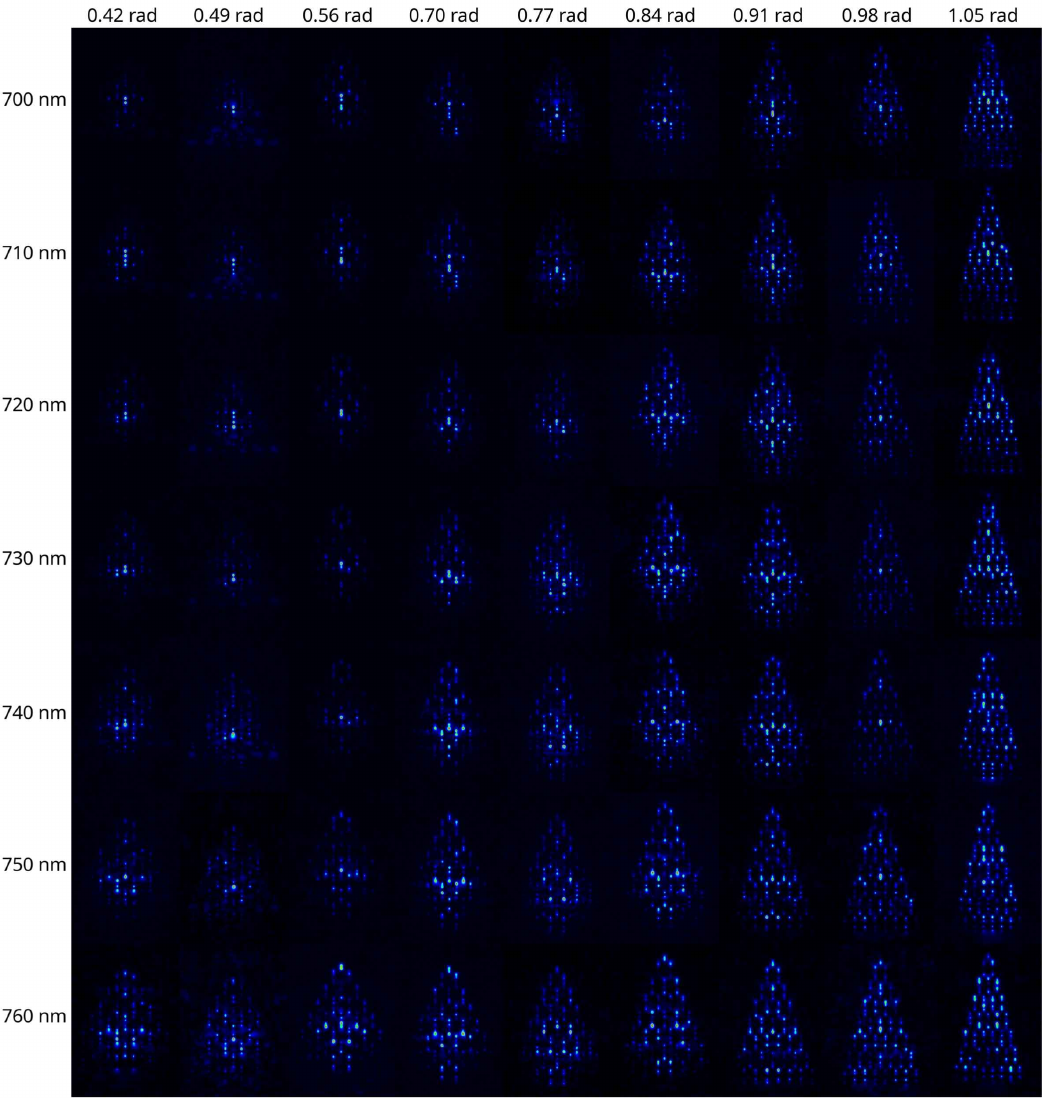}
	\caption{ 
        Supercontinuum central bulk excitation versus excitation wavelength and lattice angle $\theta$.
        }
        \label{fig:SFSCbulk}
\end{figure*}

We first observe in Fig.~\ref{fig:SFSCbulk} a compilation of results after exciting a central bulk site of our strained graphene-like lattices, for 9 different lattice angles and 7 different input wavelengths. We observe a clear localization trend around the optimized magic angle regime ($\theta\sim0.56$ rad and $730$ nm), which is well defined for shorter excitation wavelengths and smaller angles. Both tendencies diminish the coupling between $p$-mode waveguides and, therefore, makes even clearer the ABF regime as next-nearest neighbor couplings become negligible. For larger excitation wavelengths and more vertically oriented lattices (larger $\theta$), we observe an enhanced transport/diffraction through the lattice. This compilation shows clearly a phase transition from a magic angle and an all bands flat regime to the regime of mostly dispersive transport where the dispersive properties of the bands are manifested. The extracted data of the IPR for every image is shown in the main text.

\begin{figure*}[h]
	\centering
	\includegraphics[width=0.49\textwidth]{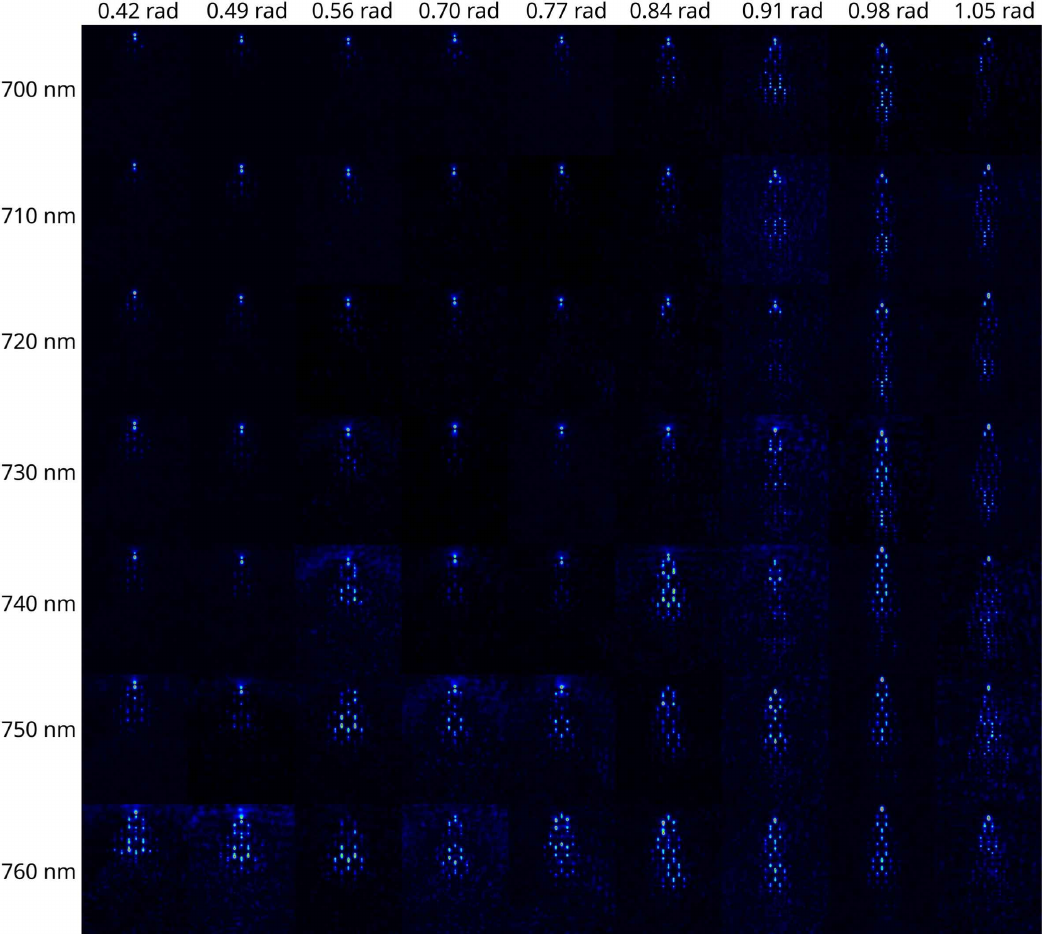}
	\caption{ 
        Supercontinuum corner-site excitation versus excitation wavelength and lattice angle $\theta$.
        }
        \label{fig:SFSCcorner}
\end{figure*}

A corner-site excitation compilation of the strained graphene-like lattices, using a SC laser source, is shown in Fig.~\ref{fig:SFSCcorner}. We observe strong and robust localized profiles over a broader set of parameters compared to its bulk-excitation counterpart, due to the strong influence of the corner topological state which is isolated from the bands around $\delta k_z \sim 0$ (rad/cm). Specifically, for wavelengths smaller than $740$ nm and angles $\theta<0.9$ we observe a rather strong localization for a perfectly periodic lattice system. 


To conclude, a SC lattice excitation allows us to characterize the performance of the system under different excitation wavelengths. This could be useful for all-optical manipulation of photonic systems, where the the localization and transport are controlled by adjusting the excitation wavelength and the geometry (angle $\theta$).

\bibliography{refs}